\documentclass[12pt,epsf,amssymb]{article}

\usepackage{graphicx}

\begin{document}
\baselineskip 0.6cm
\newcommand{\gsim}{ \mathop{}_{\textstyle \sim}^{\textstyle >} }
\newcommand{\lsim}{ \mathop{}_{\textstyle \sim}^{\textstyle <} }
\newcommand{\vev}[1]{ \left\langle {#1} \right\rangle }
\newcommand{\bra}[1]{ \langle {#1} | }
\newcommand{\ket}[1]{ | {#1} \rangle }
\newcommand{\Dsl}{\mbox{\ooalign{\hfil/\hfil\crcr$D$}}}
\newcommand{\nequiv}{\mbox{\ooalign{\hfil/\hfil\crcr$\equiv$}}}
\newcommand{\EV}{ {\rm eV} }
\newcommand{\KEV}{ {\rm keV} }
\newcommand{\MEV}{ {\rm MeV} }
\newcommand{\GEV}{ {\rm GeV} }
\newcommand{\TEV}{ {\rm TeV} }

\def\diag{\mathop{\rm diag}\nolimits}
\def\Spin{\mathop{\rm Spin}}
\def\SO{\mathop{\rm SO}}
\def\O{\mathop{\rm O}}
\def\SU{\mathop{\rm SU}}
\def\U{\mathop{\rm U}}
\def\Sp{\mathop{\rm Sp}}
\def\SL{\mathop{\rm SL}}
\def\tr{\mathop{\rm tr}}


\begin{titlepage}

\begin{flushright}
CERN-TH/2003-059\\
UT-03-09 \\
\end{flushright}

\vskip 2cm
\begin{center}
{\large \bf  Upper Bound of the Proton Lifetime \\ in Product-Group Unification}

\vskip 1.2cm
Masahiro Ibe and T.~Watari

\vskip 0.4cm
{\it Department of Physics, University of Tokyo, \\
          Tokyo 113-0033, Japan}\\

\vskip 1.5cm
\abstract{Models of supersymmetric grand unified theories based 
on the SU(5)$_{\rm GUT} \times$U($N$)$_{\rm H}$ gauge group ($N = 2,3$)
have  a symmetry that guarantees light Higgs doublets 
and the absence  of  dimension-five proton-decay operators.  
We analysed the proton decay induced by the gauge-boson exchange 
in these models. 
Upper bounds of proton lifetime are obtained;
$\tau(p\to \pi^0e^+)\lsim 6.0\times 10^{33}$ yrs in the 
SU(5)$_{\rm GUT}\times$U(2)$_{\rm H}$ model 
and $\tau(p\to \pi^0e^+)\lsim 5.3\times 10^{35}$ yrs 
in the SU(5)$_{\rm GUT}\times$U(3)$_{\rm H}$ model.
Various  uncertainties in the predictions are also discussed.
}

\end{center}
\end{titlepage}

\baselineskip 0.55cm
\section{Introduction}
\label{sec:intro}

The supersymmetric (SUSY) grand unified theories (GUTs) are among one of
the most promising candidates of physics beyond the standard model (SM); 
this is because of their theoretical beauty and 
is also because the gauge-coupling unification is supported 
by precision experiments. 
The GUTs generically predict a proton decay through 
a gauge-boson exchange\footnote{There is a class of models 
of SUSY GUTs where the gauge-boson exchange does not induce 
the proton decay \cite{Witten85}.}.
The lifetime of the proton, however, varies very much 
from model to model; 
it is proportional to the fourth power of the gauge-boson mass. 
Thus, the proton-decay experiments are not only able to provide strong
support for the GUTs but also to select some models out of 
several candidates.

For more than twenty years, many attempts have been made 
at constructing models of SUSY GUTs. 
There are two strong hints to find realistic models. 
First, the two Higgs doublets are light, 
whereas their mass term is not forbidden 
by any of the gauge symmetries 
of the minimal supersymmetric standard model (MSSM). 
Second, the rate of proton decay through dimension-five operators 
\cite{SYW} is smaller than naturally expected \cite{MPdim5}\footnote{The
experimental lower bound of the proton lifetime through those operators
is $\tau(p\rightarrow K^+ \bar\nu) \gsim 6.7 \times 10^{32}$ yrs (90\%
C.L.) \cite{SKdim5}.}.         
Thus, the two phenomena require two small parameters;
it would then be conventional wisdom to consider that
there might be symmetries behind the small parameters.
Moreover, only one symmetry is sufficient to explain the two phenomena
because of the following reason.
If a symmetry forbids the mass term of the two Higgs doublets, 
\begin{equation}
 W \rlap{/}{\ni} H_d H_u,
\end{equation}
then the dimension-five proton-decay operators are also forbidden 
by the same symmetry, 
\begin{equation}
 W \rlap{/}{\ni} QQQL + \bar{U}\bar{E}\bar{U}\bar{D},
\end{equation}
and vice versa. 
Here, it is implicitly assumed that the quarks and leptons 
have the same charge under the symmetry when they are 
in the same SU(5) multiplet, 
and that the Yukawa couplings of quarks and leptons 
are allowed by the symmetry.

There are three classes of models that have such a
symmetry\footnote{There 
are other types of models that are not constructed as field theoretical
models in four-dimensional space-time \cite{Kawamura,HN1},
\cite{WittenG2} and \cite{IWY,WY}. Reference \cite{WittenG2} is a
string-theoretical realization of \cite{BDS,Barr,DNS} and
Refs. \cite{IWY,WY} are higher-dimensional extension of \cite{IY}.   
The proton decay is analysed in \cite{HN1,HN2} for the model
\cite{Kawamura} and in \cite{FrW} for the model \cite{WittenG2}.}. 
One uses the SU(5)$_1$ $\times$SU(5)$_2$ gauge
group \cite{BDS,Barr,DNS}, where there is an unbroken {\bf Z}$_N$
symmetry \cite{WittenG2,DNS}.
The second class consists of models based on the 
SU(5)$_{\rm GUT} \times$U($N$)$_{\rm H}$ ($N=2,3$) gauge group
\cite{Yanagida,HY,IY}, where there is an unbroken discrete 
R symmetry \cite{IY}.
The last class of models can be constructed so that the unified gauge
group is a simple group \cite{Maekawa}.
The symmetry discussed in the previous paragraph, however, cannot remain
unbroken; it is broken in such a way that the dimension-five operators
are not completely forbidden. 
The proton decay through the gauge-boson exchange is discussed 
in \cite{IKNY,MY} for these models. 

In this article, we analyse the proton decay for the second class of
models.  
The dimension-five operators are completely forbidden in these models,
and the proton decay is induced by the gauge-boson exchange.
Our analysis is based on models in \cite{IY}, which use the SU(5)$_{\rm
GUT}\times$ U($N$)$_{\rm H}$ gauge group ($N=2,3$). 
Reference \cite{FW} obtained an estimate of the proton lifetime 
in the SU(5)$_{\rm GUT} \times$U(3)$_{\rm H}$ model, adopting a number
of ansatze to make the analysis simple. 
The estimate was\footnote{The following numerical value 
is based on a re-analysis in \cite{Watari}.} 
$\tau(p \rightarrow \pi^0 e^+) \simeq$ (0.7--3)$\times 10^{34}$ yrs, 
and hence there is an intriguing possibility that the proton decay is
observed in the next generation of water \v{C}erenkov detectors.
This article presents a full analysis: both the SU(5)$_{\rm
GUT}\times$U(2)$_{\rm H}$ model and the SU(5)$_{\rm
GUT}\times$U(3)$_{\rm H}$ one are analysed without any ansatze. 
Three parameters of the models are fixed by three gauge coupling
constants of the MSSM, and the remaining parameters are left
undetermined. 
The range of these parameters is restricted when the models are
required to be in the calculable regime.
As a result, the range of the gauge-boson mass is restricted.
Thus, we obtain the {\em range}\footnote{
This procedure is the one adopted in \cite{HMY}, where
the minimal SU(5) SUSY GUT model was analysed.}   
of the lifetime rather than an {\em estimate} of it.

The organization of this article is as follows. 
First, we briefly review both the SU(5)$_{\rm GUT}\times$ U($N$)$_{\rm H}$
models ($N = 2,3$) in section \ref{sec:review}.
The range of the GUT-gauge-boson mass is determined both for the
SU(5)$_{\rm GUT}\times$U(2)$_{\rm H}$ model and for the SU(5)$_{\rm GUT}
\times$U(3)$_{\rm H}$ model in sections \ref{sec:U2} and \ref{sec:U3},
respectively. 
In particular, it is shown that the range of the mass is bounded 
from above,  which leads to an upper bound of the lifetime of the proton 
in each model. 
The upper bounds\footnote{Only a lower bound is obtained, 
e.g. in the minimal SU(5) model \cite{HMY}.} 
are, in general, predictions that can be confirmed by experiments.
Since the SUSY-particle spectrum affects the MSSM gauge coupling constants 
through threshold corrections, the upper bound of the lifetime depends
on the spectrum. 
Therefore, the upper bound is shown as a function of 
SUSY-breaking parameters in section \ref{sec:ubd} for both models.
Various uncertainties in our predictions are also discussed.
Section \ref{sec:conclusion} gives a brief summary of the results obtained 
in this article, and compares the results with predictions of other models.
We find that $\tau(p\rightarrow \pi^0e^+) \lsim 6.0 \times 10^{33}$ yrs 
in the SU(5)$_{\rm GUT} \times$U(2)$_{\rm H}$ model, and 
$\tau(p\rightarrow \pi^0e^+) \lsim 5.3 \times 10^{35}$ yrs 
in the SU(5)$_{\rm GUT} \times$U(3)$_{\rm H}$ model; 
here, we exploit the uncertainties 
in the value of the QCD coupling constant (by $\pm2\sigma$) and 
in the threshold corrections from the SUSY particles; other
uncertainties, which cannot be estimated, are not included in these
figures.  

\baselineskip 0.6cm
\section{Brief Review of Models}
\label{sec:review}
\subsection{SU(5)$_{\rm GUT}\times$U(2)$_{\rm H}$ Model}

Let us first explain a model based on the product gauge group SU(5)$_{\rm
GUT} \times$U($2$)$_{\rm H}$.
Quarks and leptons are singlets of the U(2)$_{\rm H}$ gauge group and
form three families of {\bf 5}$^*$+{\bf 10} of the SU(5)$_{\rm GUT}$.
Fields introduced to break the SU(5)$_{\rm GUT}$ symmetry are given as
follows: $X^{\alpha}_{\;\;\beta}$ $(\alpha ,\beta =1,2)$, which 
transforms as (${\bf 1}$,{\bf adj.}=${\bf 3}+{\bf 1}$) 
under the SU(5)$_{\rm GUT}\times$U(2)$_{\rm H}$ gauge group, and
$Q^{\alpha}_{\;i}$ $(i=1,...,5)+Q^{\alpha}_{\;6}$ and
$\bar{Q}^i_{\;\alpha}$ $(i=1,...,5)+\bar{Q}^6_{\;\alpha}$, which
transform as ({\bf 5}$^*$+{\bf 1},{\bf 2}) and 
({\bf 5}+{\bf 1},{\bf 2}$^*$).  
The ordinary Higgs fields  $H^i({\bf 5})$ and $\bar{H}_i({\bf 5}^*)$ 
are {\em not} introduced; 
the fields $Q^{\alpha}_6$, $\bar{Q}^6_{\alpha}$ are eventually
identified with the two Higgs doublets.
The SU(5)$_{\rm GUT}$ index is denoted by $i$, and 
the U(2)$_{\rm H}$ index by $\alpha$ or $\beta$.
The chiral superfield $X^\alpha_{\;\;\beta}$ is also written as 
$X^c (t_c)^\alpha_{\;\;\beta}$ $(c=0,1,2,3)$, where $t_a$ $(a=1,2,3)$ are 
Pauli matrices of the SU(2)$_{\rm H}$ gauge group\footnote{
The normalization condition $\tr(t_a t_b) = \delta_{ab}/2$ is understood. 
Note that the normalization of the following $t_0$ is determined in such
a way that it also satisfies $\tr(t_0t_0)=1/2$.}  and $t_0 \equiv {\bf
1}_{2 \times 2}/2$, where U(2)$_{\rm H}\simeq $ SU(2)$_{\rm
H}\times$U(1)$_{\rm H}$. 
The (mod 4)-R charge assignment of all the fields in this model 
is summarized in Table \ref{tab:u2-R}. 
This symmetry forbids both the enormous mass term\footnote{ 
The mass term of the order of the weak scale for the Higgs doublets can
be obtained through the Giudice--Masiero mechanism \cite{GM}.} $W =
Q^{\alpha}_6 \bar{Q}^6_{\alpha}$ and the dangerous dimension-five
proton-decay operators $W = {\bf 10}\cdot {\bf 10} \cdot {\bf 10} \cdot
{\bf 5}^* $. 

The most generic superpotential under the R symmetry is given by
\begin{eqnarray}
W &=&\sqrt{2} \lambda_{\rm 2H} \bar{Q}^i_{\;\alpha} X^a(t_a)^{\alpha}_{\;\beta}
                             Q^{\beta}_{\;i} 
+ \sqrt{2} \lambda_{\rm 2H}' \bar{Q}^6_{\;\alpha} X^a(t_a)^{\alpha}_{\;\beta}
                             Q^{\beta}_{\;6}   \nonumber \\
& & \!\!\!\!\! 
+ \sqrt{2} \lambda_{\rm 1H} \bar{Q}^i_{\;\alpha} X^0(t_0)^{\alpha}_{\;\beta}
                             Q^{\beta}_{\;i} 
+ \sqrt{2} \lambda_{\rm 1H}' \bar{Q}^6_{\;\alpha} X^0(t_0)^{\alpha}_{\;\beta}
                             Q^{\beta}_{\;6}  \nonumber \\
& &  \!\!\!\!\!
- \sqrt{2}\lambda_{\rm 1H}  v^2 X^\alpha_{\;\alpha}    \label{eq:super2}  \\
& & \!\!\!\!\! 
+ c_{\bf 10} {\bf 10}^{i_1 i_2} {\bf 10}^{i_3 i_4} 
                   (\bar{Q}Q)^{i_5}_{\;\; 6} 
     + c_{\bf 5^*} (\bar{Q}Q)^6_{\;\; i} \cdot {\bf 10}^{ij}\cdot {\bf 5}^*_j  
      + \cdots,
\nonumber
\end{eqnarray}
where the parameter $v$ is taken to be of the order of the GUT scale;  
$\lambda_{\rm 2H},\lambda_{\rm 2H}',\lambda_{\rm 1H}$ and 
$\lambda_{\rm 1H}'$ are dimensionless coupling constants; 
$c_{\bf 10}$ and $c_{{\bf 5}^*}$ have dimensions of $(\rm mass)^{-1}$.
Ellipses stand for neutrino-mass terms and other non-renormalizable terms. 
The fields $Q^{\alpha}_{\;i}$ and $\bar{Q}^i_{\;\alpha}$ in the
bifundamental representations acquire vacuum expectation values
(VEVs), $\vev{Q^\alpha_{\;\;i}}$ = 
$v \delta^\alpha_{\;\; i}$ and $\langle\bar{Q}^i_{\;\;\alpha}\rangle$ =
$v \delta^i_{\;\;\alpha}$, because of the first three lines in
(\ref{eq:super2}).   
Thus, the gauge group SU(5)$_{\rm GUT}\times$U(2)$_{\rm H}$ 
is broken down to that of the SM.
The first terms in both the first and second lines in
 (\ref{eq:super2}) provide mass terms for the unwanted particles.
$Q_6^\alpha$ and ${\bar Q_\alpha^6}$ are identified with the Higgs
doublets in this model, and no Higgs triplets appear in the spectrum. 
As a result, no particle other than the MSSM fields, not even a gauge
singlet of the MSSM, remains in the low-energy spectrum.
This fact not only guarantees that the gauge coupling unification 
of the MSSM is maintained, but also that the vacuum is isolated. 
The R symmetry is not broken at the GUT scale, and the $\mu$ term,
dimension-four and dimension-five proton-decay operators are forbidden
by this unbroken symmetry. 

Fine structure constants of the MSSM are given at tree level by
\begin{equation}
\frac{1}{\alpha_{3}} \equiv 
 \frac{1}{\alpha_{C}} = \frac{1}{\alpha_{\rm GUT}}, \quad \quad \quad 
\label{eq:treematchU2-3}
\end{equation}
\begin{equation}
\frac{1}{\alpha_{2}} \equiv
 \frac{1}{\alpha_{L}} =  \frac{1}{\alpha_{\rm GUT}} 
                                       + \frac{1}{\alpha_{\rm 2H}}, 
\label{eq:treematchU2-2}
\end{equation}
and 
\begin{equation}
\frac{1}{\alpha_{1}} \equiv
\frac{\frac{3}{5}}{\alpha_{Y}} = \frac{1}{\alpha_{\rm GUT}} + 
                                         \frac{\frac{3}{5}}{\alpha_{\rm 1H}},
\label{eq:treematchU2-1}
\end{equation} 
where $\alpha_{\rm GUT}$, $\alpha_{2\rm H}$ and $\alpha_{1\rm H}$ 
are fine structure constants of SU(5)$_{\rm GUT}$, 
SU(2)$_{\rm H}$ and U(1)$_{\rm H}$, respectively.
Thus, the approximate unification 
of $\alpha_{3}$, $\alpha_{2}$ and $\alpha_{1}$ 
is maintained when $\alpha_{2\rm H}$ and $\alpha_{1\rm H}$ are
sufficiently large.

Although it is true that the gauge coupling unification is no longer a
generic prediction of this model, nevertheless, it need not be a mere
coincidence. 
Here, the gauge coupling unification is a consequence of the fact that
$\alpha_{2H}$ and $\alpha_{1H}$ are relatively strongly coupled when
compared with $\alpha_{GUT}$.


\subsection{SU(5)$_{\rm GUT}\times$U(3)$_{\rm H}$ Model}

The other model is based on an SU(5)$_{\rm GUT} \times$U(3)$_{\rm H}$
gauge group, 
where U(3)$_{\rm H}\simeq$ SU(3)$_{\rm H}\times$ U(1)$_{\rm H}$.   
Under the SU(5)$_{\rm GUT}\times$U(3)$_{\rm H}$ gauge group,  the
particle content of this model is $\bar{Q}^k_{\;\; \alpha}(\bf 5+1$,$\bf
3^*)$, $Q^\alpha_{\;\; k}(\bf 5^*+1$,$\bf 3)$ and $X^\alpha_{\;\;
\beta}$({\bf 1},{\bf adj}={\bf 3}$\otimes${\bf 3}$^*$), with
$k=1,\dots,5,6$; $\alpha,\beta=1,2,3$.
In addition, there are the ordinary three families of
quarks and leptons ({\bf 5}$^*$+{\bf 10},{\bf 1}) and Higgs multiplets
$H^i$+$\bar{H}_i$ ({\bf 5}+{\bf 5}$^*$,{\bf 1}).
An R symmetry forbids both (1) and (2).
The R charges of the fields are summarized in Table
\ref{tab:u3-R4}.

The most generic superpotential under the R symmetry is given
\cite{IY} by 
\begin{eqnarray}
W &=&\sqrt{2} \lambda_{\rm 3H} \bar{Q}^i_{\;\alpha} X^a(t_a)^{\alpha}_{\;\beta}
                             Q^{\beta}_{\;i} 
+ \sqrt{2} \lambda_{\rm 3H}' \bar{Q}^6_{\;\alpha} X^a(t_a)^{\alpha}_{\;\beta}
                             Q^{\beta}_{\;6}   \nonumber \\
& & \!\!\!\!\! 
+ \sqrt{2} \lambda_{\rm 1H} \bar{Q}^i_{\;\alpha} X^0(t_0)^{\alpha}_{\;\beta}
                             Q^{\beta}_{\;i} 
+ \sqrt{2} \lambda_{\rm 1H}' \bar{Q}^6_{\;\alpha} X^0(t_0)^{\alpha}_{\;\beta}
                             Q^{\beta}_{\;6}  \nonumber \\
& &  \!\!\!\!\!
- \sqrt{2}\lambda_{\rm 1H}  v^2 X^\alpha_{\;\alpha}    \label{eq:super3}  \\
  & & \!\!\!\!\! + h' \bar{H}_i \bar{Q}^i_{\;\alpha} Q^{\alpha}_{\;6} 
      + h \bar{Q}^6_{\;\alpha} Q^{\alpha}_{\;i}H^i   \nonumber \\
  & & \!\!\!\!\! + y_{\bf 10} {\bf 10} \cdot {\bf 10} \cdot H 
      + y_{\bf 5^*} {\bf 5}^* \cdot {\bf 10} \cdot \bar{H} + \cdots,
\nonumber
\end{eqnarray}
where $t_a$ $(a=1,2,...,8)$ are Gell-Mann matrices,  
$t_0 \equiv {\bf 1}_{3 \times 3}/\sqrt{6}$, 
$y_{\bf 10}$ and $y_{\bf 5^*}$ are Yukawa coupling constants 
of the quarks and leptons, and $\lambda_{\rm 3H},\lambda_{\rm 3H}',
\lambda_{\rm 1H},\lambda_{\rm 1H}',h'$ and $h$ are dimensionless coupling 
constants. 
The first three lines of (\ref{eq:super3}) 
lead to the desirable VEV of the form 
$\vev{Q^\alpha_{\;\;i}}$ = $v \delta^\alpha_{\;\; i}$ and
$\langle\bar{Q}^i_{\;\;\alpha}\rangle$ = $v \delta^i_{\;\;\alpha}$.
Thus, the SU(5)$_{\rm GUT}\times$U(3)$_{\rm H}$ gauge group 
is broken down to that of the SM.
The mass terms of the coloured Higgs multiplets arise from 
the fourth line in (\ref{eq:super3}) in the GUT-symmetry-breaking
vacuum. 
No unwanted particle remains massless.

The fine structure constants of the SU(3)$_{\rm H}\times$U(1)$_{\rm H}$
groups must be larger than that of the SU(5)$_{\rm GUT}$. 
This is because the gauge coupling constants of the MSSM are given by
\begin{equation}
 \frac{1}{\alpha_{3}}\equiv
 \frac{1}{\alpha_{C}} = \frac{1}{\alpha_{\rm GUT}} 
                          + \frac{1}{\alpha_{\rm 3H}},
\label{eq:treematchU3-3}
\end{equation}
\begin{equation}
 \frac{1}{\alpha_{2}}\equiv
 \frac{1}{\alpha_{L}} =  \frac{1}{\alpha_{\rm GUT}}, \quad \quad \quad
\label{eq:treematchU3-2}
\end{equation}
and 
\begin{equation}
 \frac{1}{\alpha_{1}}\equiv
\frac{\frac{3}{5}}{\alpha_{Y}} = \frac{1}{\alpha_{\rm GUT}} + 
                                                 \frac{\frac{2}{5}}{\alpha_{\rm 1H}},
\label{eq:treematchU3-1}
\end{equation} 
where $\alpha_{3\rm H}$ and $\alpha_{1\rm H}$ are fine structure constants 
of SU(3)$_{\rm H}$ and U(1)$_{\rm H}$, respectively. Thus, the
approximate unification of $\alpha_{3}$, $\alpha_{2}$ and $\alpha_{1}$ is
maintained when $\alpha_{3\rm H}$ and $\alpha_{1\rm H}$ are sufficiently
large.\\

\subsection{Rough Estimate of Matching Scale}
\label{ssec:est-mg}

\begin{figure}[ht]
\begin{center}
\begin{tabular}{c}
 \includegraphics[width = .6\linewidth]{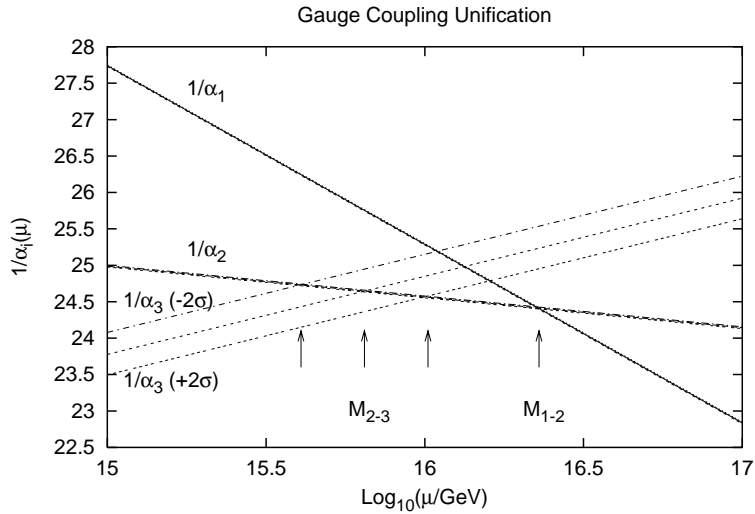} \\
\end{tabular}
\caption{Close-up view of the unification of 
the three gauge coupling constants of the MSSM. 
The fine structure constants in the $\overline{\rm DR}$ scheme 
of the U(1)$_{Y}$, SU(2)$_{L}$ and SU(3)$_{C}$ gauge interactions
are denoted by  $\alpha_{1,2,3}$, respectively.  
Three lines of $\alpha_3$ correspond to three different experimental
 inputs; the QCD coupling constants 
$\alpha_s^{\overline{\rm MS},(5)}(M_Z)=0.1132$ ($-2\sigma$), 
0.1172 (central value) and 0.1212 ($+2\sigma$) are used \cite{PDG02}. 
The 2-loop renormalization-group effects of the MSSM and the SUSY
 threshold corrections are taken into account.  
The latter corrections are those from the SUSY-particle spectrum 
determined by the mSUGRA boundary condition with $\tan \beta = 10$, 
$A_0 = 0$ GeV, ($m_0$,$m_{1/2}$) = (400 GeV, 300 GeV) and $\mu > 0$ 
(see the caption for Fig. \ref{fig:Cntr-U2-mSUGRA} for the convention on 
 the sign of $\mu$).} 
\label{fig:closeup}
\end{center}
\end{figure}

Figure \ref{fig:closeup} shows the renormalization-group evolution of 
the three gauge coupling constants of the MSSM.
The tree-level matching equations 
(\ref{eq:treematchU2-3})--(\ref{eq:treematchU2-1}) and 
(\ref{eq:treematchU3-3})--(\ref{eq:treematchU3-1}) 
suggest that the matching scale is below the energy scale $M_{2-3}$ 
in Fig. \ref{fig:closeup} 
in the SU(5)$_{\rm GUT}\times$U(2)$_{\rm H}$ model, 
and is between the two energy scales $M_{2-3}$ and $M_{1-2}$ 
in the SU(5)$_{\rm GUT}\times$U(3)$_{\rm H}$ model.
Here, $M_{2-3}$ is the energy scale where
coupling constants of the SU(2)$_{L}$ and the SU(3)$_{C}$ are
equal, and $M_{1-2}$ the one where coupling constants of the SU(2)$_{L}$
and the U(1)$_{Y}$ are equal.
In particular, the matching scale is lower than the scale $M_{1-2}$,
i.e. the conventional definition of the unification scale, in both
models. 
Thus, the decay rate of the proton is higher than the conventional estimate,
which uses $M_{1-2} \sim 2 \times 10^{16}$ GeV as the GUT-gauge-boson
mass.

\section{Gauge-Boson Mass in the SU(5)$_{\rm GUT}\times$U(2)$_{\rm H}$ Model}
\label{sec:U2}

Let us now proceed from the discussion at tree level 
to the next-to-leading-order analysis in order to draw more precise
predictions.  
To find the proton-decay rate, it is necessary to determine the mass of
the gauge boson, rather than the matching scale. 
The GUT-gauge-boson mass enters the threshold corrections to the
gauge coupling constants at the 1-loop level, and it can hence be 
discussed directly.

The analysis in this article follows the  procedure described in
\cite{HMY}. 
First, the three gauge coupling constants of the MSSM are given 
in terms of gauge coupling constants and other parameters of GUT models. 
We include 1-loop threshold corrections from the GUT-scale spectra to
the matching equations. 
Then, we constrain the parameters of GUT models 
by the three matching equations: 
three parameters are determined,  
and other parameters are left undetermined. 
The free parameters, however,  cannot be completely free when we require
that the GUT models be in a calculable regime, i.e. when a perturbation
analysis is valid. 
We determine the calculable region in the space of the free parameters 
and, as a result, the ranges of the GUT-gauge-boson masses are obtained
for the models.  

In the minimal SU(5) SUSY-GUT model, for example, 
there are four parameters.  
The three gauge coupling constants of the MSSM determine  
the coloured-Higgs mass, 
and put two independent constraints between the other three parameters.
The three parameters are the unified gauge coupling constant, 
the GUT-gauge-boson mass and a coefficient of the cubic coupling 
of the SU(5)$_{\rm GUT}$-{\bf adj.}
chiral multiplet in the superpotential \cite{HMY}.
Thus, the matching equations cannot determine the GUT-gauge-boson mass
directly. 
The cubic-coupling coefficient is chosen as the free parameter, while
the GUT-gauge-boson mass and the unified gauge coupling constant are
solved in terms of the free parameter and the MSSM gauge coupling
constants.
The free parameter, however, 
cannot be too large;
otherwise it would immediately make itself extremely large in the
renormalization-group evolution toward the ultraviolet (UV).
Thus, it is bounded from above, and its upper bound leads to the lower
bound  of the GUT-gauge-boson mass of the minimal SU(5) model \cite{HMY}. 

The SU(5)$_{\rm GUT} \times$U($N$)$_{\rm H}$ model ($N=2$ (or 3)) 
has five (or six) parameters
in the three matching equations of gauge coupling constants, as we see later. 
Thus, two (or three) parameters are left undetermined. 
The space of two (or three) free parameters is restricted by requiring the
perturbation analysis to be valid, just as in the analysis 
of the minimal SU(5) model.
As a result, the range of the GUT-gauge-boson mass is obtained.
The crucial difference between the three models is that only the lower
bound of the mass is obtained in the minimal SU(5) model, 
while the upper bound is obtained both in the SU(5)$_{\rm
GUT}\times$U(2)$_{\rm H}$ model and in the SU(5)$_{\rm
GUT}\times$U(3)$_{\rm H}$ model\footnote{
The lower bound also exists in this model.
}, as shown in the following.
The SU(5)$_{\rm GUT}\times$U(2)$_{\rm H}$ model is analysed in this
section, and the result of the SU(5)$_{\rm GUT}\times$U(3)$_{\rm H}$
model is described in section \ref{sec:U3}.  

\subsection{Parameters of the Model}
\label{ssec:U2-parameter}


The MSSM gauge coupling constants are given in terms of parameters of
the SU(5)$_{\rm GUT}\times$ U(2)$_{\rm H}$ model at the 1-loop level as 
\begin{eqnarray}
\frac{1}{\alpha_{3}}(\mu) &=& \frac{1}{\alpha_{\rm GUT}}(M)
                            \qquad \qquad \quad 
			  + \frac{3}{2\pi} \ln \left(\frac{\mu}{M}\right) 
			  + \frac{4}{2\pi} \ln \left(\frac{M_G}{M}\right),
   \label{eq:full-matchU2-3} \\
\frac{1}{\alpha_{2}}(\mu) &=& \frac{1}{\alpha_{\rm GUT}}(M) 
                          + \frac{1}{\alpha_{\rm 2H}}(M) 
                          + \frac{-1}{2\pi} \ln \left(\frac{\mu}{M}\right)
			  + \frac{6}{2\pi} \ln \left(\frac{M_G}{M}\right)
                          + \frac{4}{2\pi} 
                              \ln \left(\frac{M_{3V}}{M_{3C}}\right),
   \label{eq:full-matchU2-2} \\
\frac{1}{\alpha_{1}}(\mu) &=& \frac{1}{\alpha_{\rm GUT}}(M) 
                          + \frac{\frac{3}{5}}{\alpha_{\rm 1H}}(M) 
                          + \frac{-\frac{33}{5}}{2\pi} 
                                  \ln \left(\frac{\mu}{M}\right)
                          + \frac{10}{2\pi} \ln \left(\frac{M_G}{M}\right),
   \label{eq:full-matchU2-1}
			\end{eqnarray}
where $M$ and $\mu$ are renormalization points of the GUT
model and MSSM, respectively. 
The renormalization point $M$ is chosen to be 
above the GUT scale 
and $\mu$ to be just below the GUT scale. 
The right-hand sides consist of the tree-level contributions 
(the first and second terms) in 
Eqs. (\ref{eq:treematchU2-3})--(\ref{eq:treematchU2-1}), 1-loop
renormalization and threshold corrections  (the remaining terms). 
Gauge coupling constants are considered to be defined 
in the $\overline{\rm DR}$ scheme, and hence 
the step-function approximation is valid in the 1-loop 
threshold corrections \cite{DRB-STEPFCN}.
Various mass parameters of the model enter the equations
through the threshold corrections; 
$M_G$ is the GUT-gauge-boson mass, $M_{3V}$ and $M_{3C}$ are masses of the 
SU(2)$_{L}$-{\bf adj.} vector multiplet and chiral multiplet, 
respectively.
These mass parameters are given in terms of the parameters of the
Lagrangian (at tree level), as shown in Table \ref{tab:u2-spec}.

There are five parameters of the GUT model in the above three equations: 
$M_G$, $M_{3V}/M_{3C}$, $1/\alpha_{\rm 2H}$, $1/\alpha_{\rm 1H}$ and 
$1/\alpha_{\rm GUT}$. 
Three of them are solved in terms of the other two parameters and of the
three MSSM gauge coupling constants. 
The other two parameters are left undetermined for the moment.
We take $M_{G}$ and $M_{3V}/M_{3C}$ as the two independent
free parameters.
Then three others, namely $\alpha_{\rm GUT}(M_G)$,
$\alpha_{\rm 2H}(M_G)$ and $\alpha_{\rm 1H}(M_G)$, are determined
through Eqs. (\ref{eq:full-matchU2-3})--(\ref{eq:full-matchU2-1}) by
setting $\mu = M = M_G$.  
Another parameter of the model, $\alpha_{\rm 2H}^\lambda \equiv
(\lambda_{\rm 2H})^2/(4\pi)$, is also expressed in terms of $\alpha_{\rm
2H}(M_G)$,  
$\alpha_{\rm GUT}(M_G)$ and $M_{3V}/M_{3C}$:
\begin{equation}
 \frac{1}{\alpha_{\rm 2H}^\lambda} (M_G) = 
   \left(\frac{1}{\alpha_{\rm 2H}(M_G) + \alpha_{\rm GUT}(M_G)}\right)
   \left(\frac{M_{3V}}{M_{3C}}\right)^2.
\end{equation}

\subsection{Parameter Region of the Model
}
\label{ssec:U2-region}

Let us now determine the parameter region in the parameter space spanned by 
$M_G$ and $M_{3V}/M_{3C}$. 
We require that the perturbation analysis be valid; it is necessary that all
the coupling constants in the model be finite in the
renormalization-group evolution toward the UV, at least within the range
of spectrum of the model.  
To be more explicit, the coupling constants $\alpha_{\rm 2H}(M)$,
$\alpha_{\rm 2H}^\lambda(M)$, $\alpha_{\rm 2H}^{\lambda^{'}}(M)$, 
$\alpha_{\rm 1H}(M)$, $\alpha_{\rm 1H}^{\lambda}(M)$ and 
$\alpha_{\rm 1H}^{{\lambda^{'}}}(M)$ are required to be finite in the
renormalization-group evolution, at least while the renormalization point
$M$ is below the heaviest particle of the model. 
We use this necessary condition to determine the parameter region.
Here and hereafter, we adopt the following notation:
\begin{equation}
 \alpha_{\rm 2H}^{\lambda} \equiv 
       \frac{(\lambda_{\rm 2H})^2}{4\pi}, \quad 
 \alpha_{\rm 2H}^{\lambda^{'}} \equiv 
       \frac{(\lambda_{\rm 2H}^{'})^2}{4\pi}, \quad 
 \alpha_{\rm 1H}^{\lambda} \equiv 
       \frac{(\lambda_{\rm 1H})^2}{4\pi}, \quad 
 \alpha_{\rm 1H}^{\lambda^{'}} \equiv 
       \frac{(\lambda_{\rm 1H}^{'})^2}{4\pi}. 
\end{equation}   

First, we use the 1-loop renormalization-group equation\footnote{
The beta function of $\alpha_{\rm 2H}^{\lambda}(M)$ depends also 
on $\alpha_{\rm 2H}^{\lambda'}(M)$, $\alpha_{\rm 1H}^{\lambda}(M)$ and 
$\alpha_{\rm 1H}^{\lambda'}(M)$. 
Thus, the beta function 
cannot be calculated without the values 
of those coupling constants.
Their values, however, are not determined through the 1-loop 
matching equations (\ref{eq:full-matchU2-3})--(\ref{eq:full-matchU2-1}).
Therefore, we set them in the beta function as 0, so that
$\alpha_{\rm 2H}^{\lambda}(M)$ becomes large as slowly as possible 
in the evolution to the UV.
This makes the excluded parameter space smaller and 
makes our analysis more conservative.}
to determine the parameter region.
Renormalization-group equations of this model are listed in
appendix \ref{sec:RGE}. 
The result is shown in the left panel of Fig.~\ref{fig:U2Region}. 
The parameter region is given by the shaded region in the
$M_{G}$--$(M_{3V}/M_{3C})$ plane. 
The analysis is based on the value of $\alpha_3(\mu)$ calculated from 
$\alpha_s^{\overline{\rm MS},(5)}(M_Z)=0.1212$, i.e. 
the value larger than the central value  by $2\sigma$. 
The reason for this choice is explained shortly.

The result in the left panel is understood intuitively as follows.
First, the parameter $M_{G}$ is bounded from above (from the right of the
panel).  
It is quite a natural consequence, since it is consistent 
with the rough estimate of the matching scale in subsection
\ref{ssec:est-mg}.   
Secondly, the parameter region is bounded also from below.
It is also a natural consequence because of the following reason.
The beta function of the superpotential coupling 
$\alpha_{\rm 2H}^{\lambda}$ in Eq. (\ref{eq:RGE-U2-2HL}) implies 
that this coupling constant immediately becomes large
unless its contribution to the beta function is cancelled
by those from gauge interactions.
Thus, the parameter space with 
$\alpha_{\rm 2H} \ll \alpha_{\rm 2H}^{\lambda}$, which is almost
equivalent to $M_{3V} \ll M_{3C}$, is excluded. 

We adopt the value $\alpha_s^{\overline{\rm MS},(5)}(M_Z)=0.1212$ for
the value of the QCD coupling constant, rather than the usual central
value $\alpha_s^{\overline{\rm MS},(5)}(M_Z)=0.1172$, because this
allows $M_{2-3}$ to be larger.  
In turn, this allows the excluded region to be smaller, and hence the upper
bound for the GUT-gauge-boson mass becomes more conservative with this choice.


We further include 2-loop effects in the beta functions of the gauge
coupling constants\footnote{Note that the beta functions are
scheme-independent up to two loops for gauge coupling constants, 
while only up to one loop for coupling constants in the superpotential.}. 
The renormalization-group equations for the gauge coupling constants 
are listed in appendix \ref{sec:RGE}.
The 2-loop effects also become important at generic points of the
parameter space, because the 1-loop beta functions of the gauge coupling 
constants are accidentally small everywhere on the parameter
space\footnote{ 
At 1-loop order, the beta function of $\alpha_{\rm 2H}$ is given as a
result of cancellation between vector loop $3$T$_{G}=6$ and chiral
loop $-$T$_{R}=-8$.}.
The initial values of the 2-loop renormalization-group evolution, i.e.
values at the matching scale
$M=M_G$, are not determined for $\alpha_{\rm 2H}^{\lambda'}$,
$\alpha_{\rm 1H}^\lambda$ and $\alpha_{\rm 1H}^{\lambda'}$ from the
matching equations (\ref{eq:full-matchU2-3})--(\ref{eq:full-matchU2-1}).  
Thus, we set their values as 
\vspace{-0.2cm}
\begin{equation}
 \alpha_{\rm 2H}^{\lambda'} = \alpha_{\rm 2H}^\lambda, \qquad 
 \alpha_{\rm 1H}^{\lambda} =  \alpha_{\rm 1H}^{\lambda'} = 
 \alpha_{\rm 1H},
\label{eq:U2-initial}
\vspace{-0.2cm}
\end{equation} 
when the renormalization point $M$ is at $M_G$.
Although we should have varied also these values as free parameters, we
believe that the result of our analysis is not affected very much 
by changing these values; 
the reason is explained in appendix \ref{sec:U2-N2}.
The right panel of Fig. \ref{fig:U2Region} is 
the parameter region determined in this analysis.

The right panel of Fig. \ref{fig:U2Region} shows 
that the parameter space with $M_{3V} \gg M_{3C}$,
 i.e. $\alpha_{\rm 2H} \gg \alpha_{\rm 2H}^\lambda$, is further
excluded, and the only surviving parameter region is around the line of
$M_{3V} \simeq M_{3C}$, i.e. $\alpha_{\rm 2H}\simeq \alpha_{\rm
2H}^{\lambda}$. 
It is clear, as shown below, why this region 
and only this region survives. 
Let us neglect, for the moment, the renormalization effects 
from the SU(5)$_{\rm GUT}$ gauge interaction; 
the SU(5)$_{\rm GUT}$ gauge coupling constant is smaller 
than those of the SU(2)$_{\rm H}$ and the U(1)$_{\rm H}$
interactions. 
Then, one can see that the 2-loop part of the beta functions of 
$\alpha_{\rm 2H}$ 
and $\alpha_{\rm 1H}$
are proportional to  
$(\alpha_{\rm 2H}-\alpha_{\rm 2H}^{\lambda^{(')}})$ and 
$(\alpha_{\rm 1H}-\alpha_{\rm 1H}^{\lambda^{(')}})$.
Thus, the renormalization effects from $\alpha_{\rm 2H}$ and
$\alpha_{\rm 1H}$ are completely cancelled\footnote{
Here, we assume that Eqs. (\ref{eq:U2-initial}) are also satisfied.
}
 by $\alpha_{\rm
2H}^{\lambda^{(')}}$ and $\alpha_{\rm 1H}^{\lambda^{(')}}$  
just in that region.

The cancellation in the 2-loop beta functions is due to 
the (${\cal N}$ = 2)-SUSY structure in the GUT-symmetry-breaking
sector \cite{IWY,HY,FW}; the beta functions of gauge coupling constants
vanish at two loops and higher in perturbative expansion in the
${\cal N}$ = 2 SUSY gauge theories \cite{N2-Gauge1lp}.
Therefore, the remaining region at the 2-loop level survives 
even if higher-loop effects are included in the beta functions.

The renormalization-group evolution is determined by the 1-loop 
beta functions on the (${\cal N}$ = 2)-SUSY line 
$\alpha_{\rm 2H} \simeq \alpha_{\rm 2H}^\lambda$ 
(when the SU(5)$_{\rm GUT}$ interaction is neglected).
Therefore, we consider that the point in the parameter space indicated by
an arrow in the right panel of Fig. \ref{fig:U2Region} gives
a conservative upper bound of $M_G$.
We also consider that the upper bound so obtained 
is a good approximation of the maximum value of $M_{G}$,
although the parameter region becomes thinner and thinner 
as $M_G$ increases; see appendix \ref{sec:U2-N2} for a detailed
discussion. 
Theoretical uncertainties
in this determination of the conservative upper bound of $M_G$ are discussed in
subsection \ref{ssec:uncert2}.
A related discussion is also found in appendix \ref{sec:U2-N2}. 

Now that we know that the upper bound 
is obtained on the (${\cal N}$ = 2)-SUSY line,
it is possible to obtain the upper bound of $M_G$ without numerical
analysis. 
Indeed, the following two facts make the analysis very simple; 
$M_G$ is essentially the only free parameter on the line, 
and the 1-loop renormalization-group evolution is a good
approximation there. 

The gauge coupling constant $\alpha_{\rm 2H}$ is given at $M_G$ by 
\begin{equation}
 \frac{1}{\alpha_{\rm 2H}(M_G)}
         =\frac{-4}{2\pi}\ln\left(\frac{M_G}{M_{2-3}}\right)
\label{eq:runfromUnif} 
\end{equation}
through the matching equations 
(\ref{eq:full-matchU2-3}) and (\ref{eq:full-matchU2-2}), where a threshold
correction proportional to $\ln (M_{3V}/M_{3C})$ is neglected owing to the
${\cal N} = 2$ SUSY. 
Here, $M_{2-3}$ is defined so that $\alpha_3(M_{2-3}) = \alpha_2(M_{2-3})$.
The gauge coupling constant $\alpha_{\rm 2H}$ so determined should not
be too large because 
\begin{equation}
\frac{1}{\alpha_{\rm 2H}(M_{3V})} \simeq 
\frac{1}{\alpha_{\rm 2H}(M_G)} - 
      \frac{2}{2\pi}\ln \left(\frac{M_{3V}}{M_G}\right) \gsim 0. 
\label{eq:Upper}
\end{equation}
It follows only from the inequality in (\ref{eq:Upper})\footnote{
``3.7'' is almost independent of the SUSY threshold corrections to the
MSSM gauge coupling constants.} that 
$(2\pi)/\alpha_{\rm 2H}(M_G) \gsim 3.7$; note that $M_{3V}/M_{G}$ 
can be expressed in terms of $\alpha_{\rm 2H}$ and $\alpha_{\rm GUT}$.
Thus, the upper bound of $M_G$ is given by
\begin{equation}
 M_G \simeq e^{-\frac{2\pi}{4\alpha_{\rm 2H}(M_G)}} M_{2-3}
     \lsim e^{-\frac{3.7}{4}} M_{2-3} \simeq 0.40 \times M_{2-3}.
\label{eq:MgM23}
\end{equation}

\subsection{Uncertainties in the Upper Bound of the Gauge-Boson Mass}
\label{ssec:uncert2}

Here, we estimate uncertainties in our prediction of the upper bound
of the GUT-gauge-boson mass. 
Uncertainties arising from our analysis of the GUT model are discussed
in this subsection, while uncertainties arising from low-energy physics 
are discussed in section  
\ref{sec:ubd}. 

First, we focus on the effects from the SU(5)$_{\rm GUT}$ interaction.
They have been neglected\footnote{
They are not neglected in the numerical analysis in Fig.
\ref{fig:U2Region}. 
}
in the discussion of the previous subsection, but they do contribute to
the 2-loop beta functions; in addition, the higher-loop contributions 
from $\alpha_{\rm 2H}$ and $\alpha_{\rm 2H}^\lambda$ no longer cancel 
because the SU(5)$_{\rm GUT}$ interaction does  not preserve ${\cal 
N}$ = 2 SUSY. 
Thus, the renormalization-group evolution is changed and the
determination of the upper bound is affected. 
The SU(5)$_{\rm GUT}$ interaction contributes to the beta function 
of $\alpha_{\rm 2H}$ in Eq.  (\ref{eq:RGE-U2-2H}) by less than
10\% of the 1-loop contribution\footnote{See appendix 
\ref{sec:U2-N2} for more details.
}. 
Thus, the value of $\alpha_{\rm 2H}(M_G)$ for the upper-bound value of
$M_G$ is not changed by 10\% (see Eq. (\ref{eq:Upper})).
As a result, the upper bound of $M_G$ is not
modified by a factor of more than $e^{(2\pi)/(4\alpha_{\rm 2H}(M_G))\times
(\pm 10\%)}\sim 10^{\pm 0.04}$.

Second, the perturbative expansion would not converge when 
the 't Hooft coupling $2\alpha_{\rm 2H}/(4\pi)$ exceeds unity. 
It is impossible to extract any definite statement on the
renormali-zation-group evolution 
when the perturbative expansion is not valid. 
However, most of the renormalization-group evolution 
is in the perturbative regime, i.e. $(4\pi/(2\alpha_{\rm 2H}))\gsim 1$, 
since we know that $(4\pi/(2 \alpha_{\rm 2H}))(M_G)\simeq 3.7$ 
for the upper-bound value of $M_G$. 
Thus, we consider that the perturbation analysis in the previous
subsection is fairly reliable.

Third, non-perturbative contributions are also expected in the beta
functions, and they might not be neglected since 
the gauge coupling constants are relatively large in this model.
They\footnote{We thank
Tohru Eguchi for bringing this issue to our attention.} 
are expected to be of the form \cite{Arkani-Hamed:1997mj}
\begin{equation}
 \left(1-\frac{{\rm T}_G}{2\pi}\alpha\right)
 \frac{\partial}{\partial \ln \mu}\left(\frac{1}{\alpha}(\mu)\right)
  = \frac{3{\rm T}_G - {\rm T}_R}{2\pi} + 
    \sum_{n=1}^{\infty} c_n \left(g^{-2{\rm T}_G}
                        e^{-\frac{2\pi}{\alpha}}\right)^n,
\label{eq:NPbeta}
\end{equation}
where $c_n$'s are numerical factors of the order of unity.
Each contribution comes from $n$-instantons.
Here, we neglected perturbative and non-perturbative contributions 
through wave-function renormalizations of hypermultiplets.
This is because hypermultiplets of 
${\cal N}$ = 2 SUSY gauge theories are protected from any radiative
corrections \cite{N2-Hyptree}.  
We see that the non-perturbative effects given above are not significant
when the renormalization point is around the GUT scale, since 
\begin{equation}
(g_{\rm 2H})^{-4} e^{-\frac{2\pi}{\alpha_{\rm 2H}}} 
   \simeq 5 \times 10^{-5} \ll 1. 
\end{equation}

So far, the analysis is based on a renormalizable theory.
However, the Yukawa couplings of quarks and charged
leptons are given by non-renormalizable operators in (\ref{eq:super2}).
Another non-renormalizable operator is also necessary to account for 
the fact that the Yukawa coupling constants of strange quark and muon
are not unified in a simple manner.
Those operators, however, affect the gauge coupling constants 
through renormalization group only at higher-loop levels. 
Moreover, they are not relevant to the renormalization-group flow 
(say, in the sense of Wilsonian renormalization group) 
except near the cut-off scale $M_*$. 
These are the reasons why we neglected the effects of those operators. 

There may be, however, a non-renormalizable operator
\begin{equation}
W = 2 \tr \left(\left(\frac{1}{4 g^2}+ 
c \frac{\vev{\bar{Q}Q}}{M_*^2}\right)
{\cal W}^\alpha{\cal W}_\alpha \right),
\label{eq:non-ren}
\end{equation}
which directly modifies the matching equations 
of the gauge coupling constants at tree level.
Exactly the same analysis as in subsections \ref{ssec:U2-parameter} 
and \ref{ssec:U2-region} tells us that  
the upper bound of $M_G$ given in Eq. (19) is modified\footnote{ 
Contributions to
Eqs. (\ref{eq:full-matchU2-3})--(\ref{eq:full-matchU2-1}) are, 
for example, $c \times 16\pi \times 0.024 \simeq 1.2 \times c$   
at the point in the parameter space indicated by an arrow in
Fig. \ref{fig:U2Region}.} into 
\begin{equation}
M_G \lsim 0.40 \times 10^{-0.82\times c} \times M_{2-3},
\end{equation}
as long as $c\gsim (-3.7/(6\pi)+14/9\times (\epsilon_g/\alpha_{\rm
GUT}))/1.2 \simeq -0.16+1.3\times(\epsilon_g/\alpha_{\rm GUT})$;
here, $\epsilon_g \equiv ((g_3 - g_1)/g_1)(M_{1-2}) \simeq$ 
$-$(0.03--0.01).  

\section{Gauge-Boson Mass in the SU(5)$_{\rm GUT}\times$U(3)$_{\rm H}$ Model}
\label{sec:U3}

\subsection{Parameter Region of the Model}

The same analysis as in section \ref{sec:U2} is performed 
for the SU(5)$_{\rm GUT}\times$U(3)$_{\rm H}$ model.
The 1-loop matching equations in this model, which are quite similar to 
Eqs. (\ref{eq:full-matchU2-3})--(\ref{eq:full-matchU2-1}), are found in
\cite{FW}. 
The particle spectrum around the GUT scale, which comes into 
the threshold corrections, is summarized 
in Table \ref{tab:U3-spec}. 

There are six parameters in the matching equations: 
$M_G$, $M_{8V}/M_{8C}$, $1/\alpha_{\rm GUT}$, 
$1/\alpha_{\rm 3H}$, and $1/\alpha_{\rm 1H}$, just as 
in the SU(5)$_{\rm GUT}\times$U(2)$_{\rm H}$ model, and
$M_{H_c}M_{H_{\bar{c}}}/M_G^2$. 
Three of them are fixed through the matching equations, and the other
three are left undetermined.
We take $M_G$, $M_{8V}/M_{8C}$ and $M_{H_c}M_{H_{\bar{c}}}/M_G^2$ 
as the three free parameters.
The space of these parameter is restricted 
by requiring that all the coupling constants $\alpha_{\rm 3H}(M)$, 
$\alpha_{\rm 3H}^{\lambda(')}$, $\alpha_{\rm 1H}(M)$ and  
$\alpha_{\rm 1H}^{\lambda(')}$ be finite 
in the renormalization-group evolution toward the UV, at least within the
range of the spectrum.

\begin{figure}[ht]
\begin{center}
 \includegraphics[width=.6\linewidth,height=.4\linewidth]{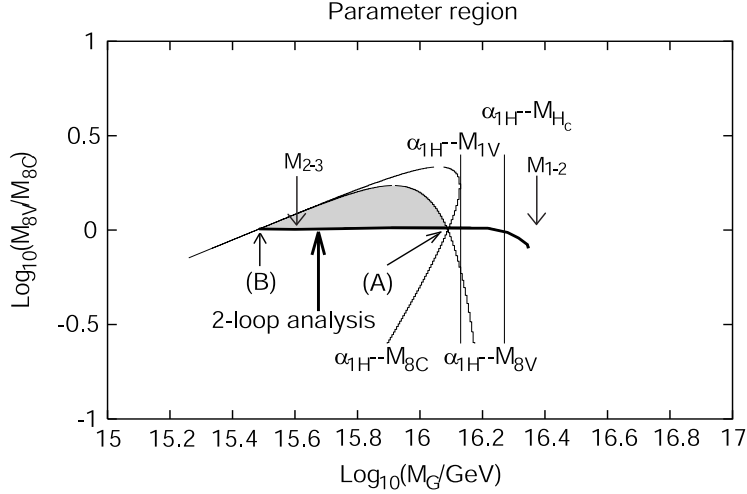} 
\caption{Parameter region of the SU(5)$_{\rm GUT}\times$U(3)$_{\rm H}$
 model. The parameter space of the model is spanned by three independent
 parameters: $M_G$, $M_{8V}/M_{8C}$ and $(M_{H_c}M_{H_{\bar{c}}})/M_G^2$.
The figure is the
$\sqrt{(M_{H_c}M_{H_{\bar{c}}})/M_G^2} = 10^{0.3}$ cross section 
of the parameter space. 
We require that all the coupling constants remain finite 
under the renormalization group, while the renormalization point 
is below the heaviest particle of the model. 
This condition is satisfied in the shaded region when 
the 1-loop renormalization group is used.
Thin curves and lines labelled  ``(gauge coupling)-mass'' are 
lines where the corresponding gauge coupling constants 
become infinite at the corresponding mass scales. 
After two loop effects are included in the beta functions of the gauge
 coupling constants, the remaining allowed parameter region is only on
 the thick curve labelled 2-loop analysis. 
Points (A) and (B) denote the upper and the lower bound of the
 gauge-boson mass $M_G$, respectively, for fixed
 $\sqrt{(M_{H_c}M_{H_{\bar{c}}})/M_G^2} = 10^{0.3}$.   
The upper bound of $M_G$ in the model is obtained as the maximum value
 $M_G$ takes at (A) as 
 $\sqrt{(M_{H_c}M_{H_{\bar{c}}})/M_G^2}$  changes.        
One also sees immediately that the lower bound at (B) is so low
 that it is of no physical importance.
$M_{1-2}$ ($M_{2-3}$) indicates the unification point between
 $1/\alpha_{1}$ and $1/\alpha_2$ ( $1/\alpha_{2}$ and
 $1/\alpha_3(-2\sigma)$ ) (see Fig. \ref{fig:closeup} for more
 details). 
Note that ($M_{G}$ at (A)) $< 10^{16.13}$ GeV 
$\simeq$ (($0.60 \simeq 10^{-0.22}$) $\times$ 
($M_{1-2} \simeq10^{16.35}$ GeV)).
The QCD coupling constant 
$\alpha_s^{\overline{\rm MS},(5)}(M_Z)=0.1132$ is used.
The effects from a non-renormalizable operator that corresponds to 
 (\ref{eq:non-ren}) in this model are  not included here.  
} 
\label{fig:U3Region}
\end{center}
\end{figure}

The parameter region is shown in Fig. \ref{fig:U3Region};  
only the $\sqrt{(M_{H_c}M_{H_{\bar c}}/M_G^2)}=10^{0.3}$ cross section
is described, and hence the region is described in the
$M_G$--$(M_{8V}/M_{8C})$ plane. 
This analysis is based on the value of $\alpha_3(\mu)$ that is
calculated from $\alpha^{\overline{\rm MS},(5)}_s(M_Z)=0.1132$, i.e. a
value smaller than the central value by 2$\sigma$. 
This is because it makes our analysis more conservative.
The parameter region obtained by the 1-loop renormalization group is
shown as the shaded area in Fig. \ref{fig:U3Region}.
The region is bounded from the right and from the left, which is again
consistent with the rough estimate of the matching scale given in
subsection \ref{ssec:est-mg}. 
The region is also bounded from below just for the same reason
as in subsection \ref{ssec:U2-region}.
The parameter region distant from the (${\cal N}$ = 2)-SUSY
line\footnote{ 
The ${\cal N}$ = 2 SUSY is enhanced in the GUT-symmetry-breaking sector
when 
\begin{equation}
 g_{\rm 1H} \simeq  \lambda_{\rm 1H}( \sim \lambda_{\rm 1H}'), \qquad \qquad
 g_{\rm 3H} \simeq  \lambda_{\rm 3H}( \sim \lambda_{\rm 3H}'),  
\end{equation}
are satisfied and $h$, $h^{'}$, $\alpha_{\rm GUT}$ are neglected
\cite{HY,IWY,FW}. 
}
$M_{8V}\simeq M_{8C}$ is excluded when 2-loop effects are
included\footnote{
We set the initial values ($M=M_G$) of coupling constants that are not
determined by the matching equation as follows:
\vspace{-0.3cm}
\begin{equation}
h=h', \qquad  
\alpha_{\rm 3H}^{\lambda} = \alpha_{\rm 3H}^{\lambda^{'}}, \qquad 
\alpha_{\rm 1H}^{\lambda} = \alpha_{\rm 1H}^{\lambda^{'}} = \alpha_{\rm 1H}.
\vspace{-0.3cm}
\end{equation}
This choice makes the renormalization-group evolution the most stable.
} 
in the beta functions of the gauge coupling constants; the 2-loop
contributions have significant effects compared with 1-loop effects,
because the 1-loop beta function of $\alpha_{\rm 3H}$ accidentally
vanishes.  
The 1-loop renormalization-group evolution is reliable on the (${\cal N}$ =
2)-SUSY line, the thick line in Fig.~\ref{fig:U3Region}, 
and hence the points indicated by (A) and (B) give 
the upper and the lower bound of $M_G$,
respectively, in the $\sqrt{(M_{H_c}M_{H_{\bar c}}/M_G^2)}=10^{0.3}$ 
cross section. 
The upper and the lower bound of $M_G$ of the model are the
maximum and the minimum value that $M_G$ takes at (A) and (B),
respectively, when the remaining parameter
$(M_{H_c}M_{H_{\bar{c}}}/M_G^2)$ is varied.   
Since it is evident from the figure that the lower bound of $M_G$ leads
to too fast a proton decay,
we focus in the following only on the upper bound of $M_G$.
 
\subsection{Uncertainties in the Upper Bound of the Gauge-Boson Mass}
\label{ssec:uncert3}

In this subsection, we estimate uncertainties in the prediction of the 
upper bound of the GUT-gauge-boson mass obtained in the previous
subsection.  
The uncertainties that originate from low-energy physics, however,
are discussed in section \ref{sec:ubd}.

First, we discuss the effects of the interactions that violate the
${\cal N}$ = 2 SUSY. 
The SU(5)$_{\rm GUT}$ gauge interaction and the cubic couplings in the
fourth line of (\ref{eq:super3}) are the sources of the violation of the
${\cal N}$ = 2 SUSY.  
Those interactions change the 1-loop-exact evolution of ${\cal N}$ = 2
SUSY gauge theories.
The change in the upper bound of $M_G$ comes\footnote{
We confirmed that the change in the evolution of $\alpha_{\rm 3H}$ is
not so significant as to make the finiteness of $\alpha_{\rm 3H}$ a
more severe condition than that of $\alpha_{\rm 1H}$.
} from the change in the evolution of $\alpha_{\rm 1H}$,
because the upper bound was determined by the running of $\alpha_{\rm
1H}$ in the absence of (${\cal N}$ = 2)-SUSY breaking.  
The beta function of $\alpha_{\rm 1H}$ is changed at most by a few
per cent\footnote{ 
This estimate comes from the ratio between the 1-loop contribution and 
the SU(5)$_{\rm GUT}$ contribution at two loops. 
Note also that the $\alpha_{h}\equiv  h^2/(4\pi)$ contribution has a
sign opposite to that of the SU(5)$_{\rm GUT}$ contribution.
}, which leads to the change  of the upper bound of $M_G$ by a factor of
at most $10^{\pm 0.01}$.   


Second, one can see from the matching equations \cite{FW} of this model
that the gauge coupling constants $\alpha_{\rm 3H}$ and $\alpha_{\rm
1H}$ are not so large as to invalidate the perturbative expansion when
the coloured-Higgs particles are moderately heavier than the
GUT-gauge-boson mass; only one threshold correction from the
coloured-Higgs particles is sufficient to keep both coupling
constants within the perturbative regime.
Non-perturbative effects are not important at all in such a region.

Finally, a non-renormalizable operator that corresponds 
to (\ref{eq:non-ren}) may also exist in this model. 
Such an operator, if it exists, contributes to the matching equations at  
tree level. 
In its presence, we can perform exactly the same analysis as in the
previous subsection. The result of this analysis is
presented in section \ref{sec:conclusion}.

\section{Conservative Upper Bound of Proton Lifetime}
\label{sec:ubd}

The analysis in sections \ref{sec:U2} and \ref{sec:U3} presented the way
of extracting the upper bound of the GUT-gauge-boson mass for both models.   
The lifetime of the proton through the GUT-gauge-boson exchange is given
\cite{HMY} in terms of $M_G$ as  
\begin{equation}
\tau(p \rightarrow \pi^0 e^+) \simeq 1.0 \times 10^{35} \times
\left(\frac{0.015 {\rm GeV}^3}{\alpha_{\rm H}}\right)^2
\left(\frac{2.5}{A_R}\right)^2
\left(\frac{1}{25 \alpha_{\rm GUT}(M_G)}\right)^2
\left(\frac{M_G}{10^{16}\GEV}\right)^4 {\rm yrs},
\label{eq:plife}
\end{equation}
where $\alpha_{\rm H}$ is a hadron matrix element\footnote{The hadron matrix
element $\alpha_{\rm H}$ is defined by 
$\bra{\rm vac.}(u_Rd_R)u_L\ket{p(\vec{\bf k})}= \alpha_{\rm H} u(\vec{\bf k})$.
This is related to another matrix element $W$ ($\simeq - 0.15 \pm 0.02$
GeV$^2$) through   
\vspace{-0.2cm}
\begin{equation}
 \frac{\alpha_{\rm H}}{\sqrt{2}f_\pi} = -W,
\vspace{-0.2cm}
\end{equation}
where $W$ is defined by $\displaystyle{\lim_{\vec{\bf p}\rightarrow {\bf
0}}}\bra{\pi^0(\vec{\bf p})}(u_Rd_R)u_L\ket{p(\vec{\bf k})}= W u(\vec{\bf k})$
, and $f_\pi$ is the pion decay constant ($2f_{\pi}=130\pm 5$ MeV
\cite{PDG02}).}  
calculated with lattice quenched QCD \cite{JLQCD} 
($\alpha_{\rm H} = -0.015 \pm 0.001$ GeV$^3$)
renormalized at 2.3 GeV and $A_R \simeq 2.5$ a renormalization factor 
of the dimension-six proton-decay operators \cite{Hisano:2000dg}; 
$A_R$ consists of a short-distance part, $A_R^{(\rm SD)} \simeq 2.1$,
which comes from the renormalization between the GUT scale and the
electroweak scale, and a long-distance part\footnote{The numerical
coefficient of the formula of the  lifetime adopted in \cite{FW} is
different from the one in Eq.~(\ref{eq:plife}) in this article.  
This is because the formula in \cite{FW} is based implicitly on  $A_R
\simeq 3.6$ in \cite{IN}, whose value is the effect of renormalization
between the GUT scale and 1 GeV.      
It was therefore incorrect, in \cite{FW}, to use at the same time $A_R
\simeq 3.6$ renormalized at 1 GeV and the hadron matrix element in
\cite{JLQCD} renormalized at 2.3 GeV.},   
$A_R^{(\rm LD)} \simeq 1.2$, from the renormalization between the
electroweak scale and 2.3 GeV ($A_R=A_R^{(\rm SD)}\cdot A_R^{(\rm
LD)}$). 
We note the expression of $A_R^{(\rm SD)}$ \cite{IN} for later
convenience: 
\begin{equation}
 A_R^{(\rm SD)} = 
\left(\frac{\alpha_C(M_Z)}{\alpha_C(M_G)}\right)^{\frac{4/3}{b_3=3}}
\left(\frac{\alpha_L(M_Z)}{\alpha_L(M_G)}\right)^{\frac{3/2}{b_2=-1}}
\left(\frac{\alpha_Y(M_Z)}{\alpha_Y(M_G)}\right)^{\frac{23/30}{b_1=-33/5}},
\label{eq:AR_SD}
\end{equation}
where $b_i$ ($i=1,2,3$) are coefficients of the 1-loop beta functions of
the three gauge coupling constants of the MSSM. 
The renormalization from Yukawa coupling constants is omitted because
its effect is negligible. 

Threshold corrections from SUSY particles are of the same order as 
those from the particles around the GUT scale. 
The 2-loop effects in the renormalization-group evolution 
between the electroweak scale and the GUT scale are also of the same order.
Therefore, the two above effects should be taken into consideration 
in deriving predictions on the GUT-gauge-boson mass (and hence on the
lifetime of the proton).
This implies, in particular, that the predictions depend on
the spectrum of SUSY particles. 
We present the predictions on the upper bound of the lifetime of the
proton as a function of SUSY-breaking parameters of the mSUGRA boundary
condition in subsection  \ref{ssec:mgugra}.
Predictions can be obtained also for other SUSY-particle spectra such as
that of gauge-mediated SUSY breaking (subsection
\ref{ssec:othermediation}). 
Subsection \ref{ssec:vect} discusses how the predictions are
changed when there are vector-like SU(5)$_{\rm GUT}$-multiplets 
at a scale below the GUT scale.

\subsection{mSUGRA SUSY Threshold Corrections}
\label{ssec:mgugra}

Let us first consider the SUSY-particle spectrum determined by the
mSUGRA boundary condition. 
This spectrum and the MSSM gauge coupling constants 
in the $\overline{{\rm DR}}$ scheme  
are calculated in an iterative procedure. 
We use the {\em SOFTSUSY1.7} code \cite{SOFTSUSY} for this purpose.
These coupling constants are evolved up to the GUT scale 
through the 2-loop renormalization group. 
They are used as input in the left-hand sides of, say, 
Eqs. (\ref{eq:full-matchU2-3})--(\ref{eq:full-matchU2-1}), 
to obtain a prediction of the upper bound of the GUT-gauge-boson mass.
The universal scalar mass $m_0$ and the universal gaugino mass $m_{1/2}$
are varied, while  we fix the other parameters of the mSUGRA boundary 
condition\footnote{  
This is because changes in these parameters did not change the result at
all as in \cite{FW}.}; $\tan \beta=10.0$, $A_0= 0$ GeV, and the sign of
the $\mu$ parameter is the standard one.

\begin{figure}
\begin{center}
\begin{tabular}{cc}
 \includegraphics[width=.4\linewidth]{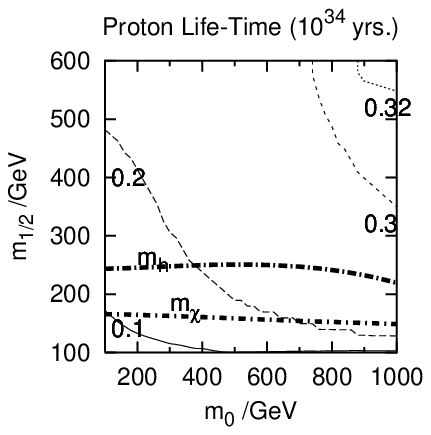}  & 
 \includegraphics[width =.35\linewidth]{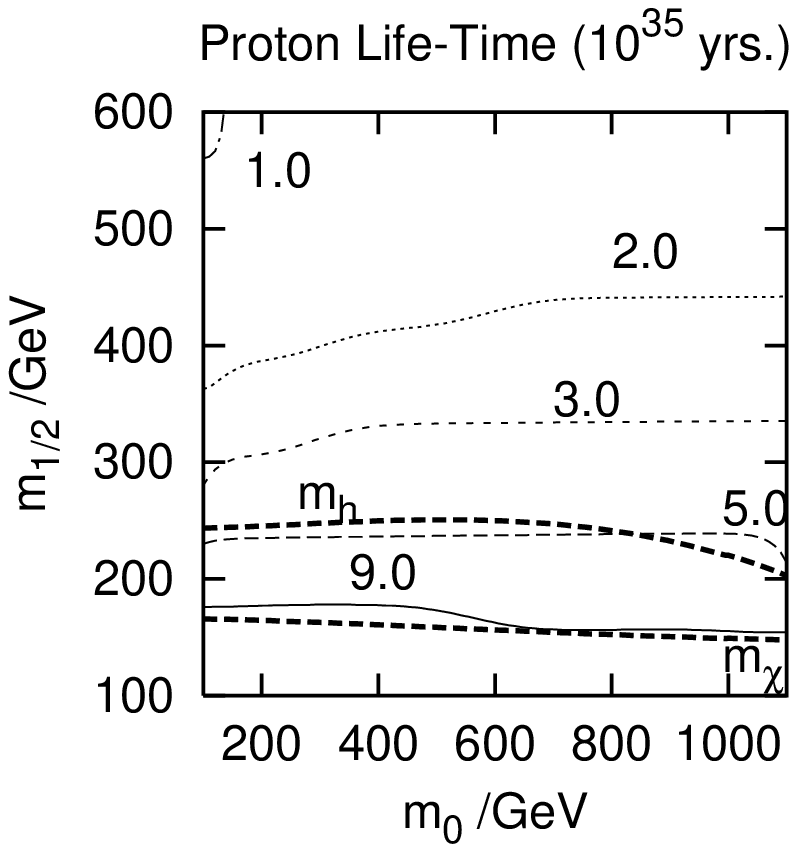}  \\          
\end{tabular}
\caption{Contour plots of the upper bound of the proton lifetime on the
mSUGRA parameter space. 
The left panel is the prediction 
of the SU(5)$_{\rm GUT}\times$U(2)$_{\rm H}$ model and 
the right one that of the SU(5)$_{\rm GUT}\times$U(3)$_{\rm H}$ model.
The upper bound changes as the universal scalar mass $m_0$ 
and the universal gaugino mass $m_{1/2}$ are varied (other mSUGRA
 parameters are  fixed at $\tan \beta = 10$, $A_0 =0.0$). 
The $\mu$ parameter is chosen to be positive, when the constraint from the
branching ratio of the $b \rightarrow s \gamma$ process is less severe. 
The upper bound of the lifetimes varies as 
(1.4--3.2)$\times 10^{33}$ yrs 
in the SU(5)$_{\rm GUT}\times$U(2)$_{\rm H}$ model, 
where the QCD coupling constant 
$\alpha_s^{\overline{\rm MS},(5)}(M_Z)=0.1212$ is used.
The upper bound varies as (1--5)$\times 10^{35}$ yrs 
in the SU(5)$_{\rm GUT}\times$U(3)$_{\rm H}$ model, 
where the QCD coupling constant 
$\alpha_s^{\overline{\rm MS},(5)}(M_Z)= 0.1132$ is used.
In both panels, the effects from non-renormalizable operators 
are not included. 
The thick curves labelled $m_h$ and $m_{\chi}$ are the bounds on the
 mSUGRA parameter space from the LEP II experiment in search of 
the lightest Higgs ($m_h \geq 114$ GeV, 95\% C.L. ) \cite{LHiggs} and   
the lightest chargino ($m_\chi \geq 103.5$ GeV, 95\% C.L.) 
\cite{chargino}. 
These curves are obtained by using the {\em SOFTSUSY1.7} code
\cite{SOFTSUSY}. 
The excluded region changes when other codes are used; 
lower bound of $m_{1/2}$ for fixed $m_0$ can be higher by about 100 GeV. 
The code we adopt yields the largest pole mass of 
the lightest Higgs scalar 
among various codes available \cite{Allanach:2003jw}, 
and hence the excluded region is the smallest. 
}
\label{fig:Cntr-U2-mSUGRA}
\end{center}
\end{figure}

The left panel of Fig. \ref{fig:Cntr-U2-mSUGRA} is a contour plot on the
$m_0$--$m_{1/2}$ plane, describing the upper bound of the proton lifetime 
in the SU(5)$_{\rm GUT}\times$U(2)$_{\rm H}$ model, where we set the
unknown coefficient $c$ of the non-renormalizable operator (22) to zero. 
The QCD coupling constant $\alpha_s^{\rm \overline{MS},(5)}(M_Z)=0.1212$
is used, so that the upper bound becomes more conservative. 
One can see that this upper bound ranges over (1.4--3.2)$\times10^{33}$
yrs. 
Notice that the ($m_0$, $m_{1/2}$) dependence of the proton lifetime
arises almost only through the variation of $M_{2-3}$ (see
Eq. (\ref{eq:MgM23})).    
Indeed, the contours of the upper bound of the lifetime  
in the left panel of Fig. \ref{fig:Cntr-U2-mSUGRA} behave in the same
way as those of $M_{2-3}$ in the upper-left panel of 
Fig. \ref{fig:mSUGRA-GMSB}. 

It is now easy to see how much the prediction is changed when we adopt 
the central value of the QCD coupling constant, $\alpha_s^{\rm
\overline{MS},(5)}(M_Z)=0.1172$. 
Since the choice of the QCD coupling constant directly changes $M_{2-3}$, 
it severely affects the upper bound of $M_G$ in this model.
$M_{2-3}$ is decreased by a factor 
$e^{-\frac{2\pi}{b_3-b_2}(\frac{1}{0.1172}-\frac{1}{0.1212})}$,  
and the lifetime is shortened by a factor $e^{-2\pi\times 0.28 } \simeq 0.2$.
We confirmed that the upper bound of the lifetime does not exceed 
$1 \times 10^{33}$ yrs even when the SUSY-breaking parameters $m_0$ 
and $m_{1/2}$ are varied up to 2000 GeV if we adopt the central value of
the QCD coupling constant.

The hadron matrix element $\alpha_{\rm H}$ in \cite{JLQCD}, which has a
statistical error $\alpha_{\rm H} = -0.015 \pm 0.001 {\rm GeV^3}$, does
not include a systematic error (e.g. an error due to the quenched
approximation).   
Reference \cite{Raby} estimates that the systematic error is about 50\%,
which leads to an uncertainty in the lifetime of a  factor of 2.   

Therefore, the conservative upper bound is roughly 
$\tau \lsim 6 \times 10^{33}$ yrs, where we exploit 
the uncertainties in the SUSY threshold corrections, 
in the value of the QCD coupling constant 
and in the hadron matrix element.
Thus, the prediction does not contradicts 
the experimental lower bound from Super-Kamiokande 
$\tau(p\rightarrow \pi^0 e^+) \gsim 5 \times 10^{33}$ yrs (90\% C.L.)
\cite{TITAND} at this moment\footnote{
The lifetime listed in \cite{PDG02} is $\tau(p\rightarrow \pi^0 e^+)
\gsim 1.6 \times 10^{33}$ yrs (90\% C.L.), based on a paper
\cite{SKdim6} published in 1998.}, 
yet the large portion of the parameter region is already excluded. 
Moreover, one can expect that 
the uncertainties originating from low-energy physics 
will be reduced in the future. 
Thus, further accumulation of data in Super-Kamiokande and the
next generation of water-\v{C}erenkov detectors  
will be sure either to exclude this model 
without the non-renormalizable operator
(\ref{eq:non-ren}), or to detect the proton decay.  

Now, we move to consider the SU(5)$_{\rm GUT}\times$U(3)$_{\rm H}$ model. 
The right panel of Fig. \ref{fig:Cntr-U2-mSUGRA} is a contour plot on the
$m_0$--$m_{1/2}$ plane, describing the upper bound of the proton lifetime 
in the SU(5)$_{\rm GUT}\times$U(3)$_{\rm H}$ model.
The QCD coupling constant $\alpha_s^{\rm \overline{MS},(5)}(M_Z)=0.1132$
is used.  
The upper bound of the proton lifetime 
ranges over (1--5)$\times 10^{35}$ yrs 
on the mSUGRA parameter space that is not excluded 
by the LEP II bound on the lightest-Higgs-boson mass.  

Let us now see how much the above prediction is changed 
by uncertainties related to the QCD. 
First, the following observation is important in discussing 
the effect from the uncertainty in the value 
of the QCD coupling constant.  
The behaviour of the contours of the upper bound
of the lifetime, and hence of the GUT-gauge-boson mass has, 
in this model, strong correlations 
with that of $M_{1-2}$ presented in the upper-right
panel of Fig.~\ref{fig:mSUGRA-GMSB}.  
We find an empirical relation 
\begin{equation}
 M_G \lsim 0.60 \times M_{1-2}.
\label{eq:MgM12}
\end{equation}
Thus, the upper bound does not depend on the value of the QCD coupling
constant very much, since $M_{1-2}$ is not affected very much.
Second, the uncertainty in the hadron form factor is common 
to both models.
Therefore, the most conservative upper bound of the proton lifetime 
is roughly $\tau \lsim 10^{36}$ yrs in this model. 
In particular, the proton decay might not be within the reach of
the next generation of experiments. 

\subsection{Threshold Corrections from Various SUSY-Particle Spectrum}
\label{ssec:othermediation}

Gauge-mediated SUSY breaking (GMSB) is one of the highly motivated models of
SUSY breaking.
The spectrum of the SUSY particles is different from that of the mSUGRA
SUSY breaking, and moreover, there are extra SU(5)$_{\rm GUT}$-charged
particles as messengers.
Thus, the predictions on the proton lifetime are different from 
those in the case of mSUGRA SUSY breaking.
We discuss the effects of the difference in the SUSY-particle spectra
in this subsection. 
A possible change of predictions due to the existence of extra
particles is discussed in the next subsection.

The ranges of the GUT-gauge-boson masses is different for different
SUSY-particle spectra, yet the difference only arises from the
difference in the two energy scales $M_{2-3}$ and $M_{1-2}$: 
the energy scale where the SU(2)$_L$ and SU(3)$_C$ coupling constants become
the same and where those of U(1)$_Y$ and SU(2)$_L$ become the same, 
respectively.
The upper bound of $M_G$ is given in terms of $M_{2-3}$ through
Eq. (\ref{eq:MgM23}) in the SU(5)$_{\rm GUT}\times$U(2)$_{\rm H}$ model,
and  in terms of $M_{1-2}$ through Eq. (\ref{eq:MgM12}) in the
SU(5)$_{\rm GUT}\times$U(3)$_{\rm H}$ model. 

\begin{figure}[ht]
 \begin{center}
  \begin{tabular}{cc}
   \resizebox{7cm}{!}{\includegraphics{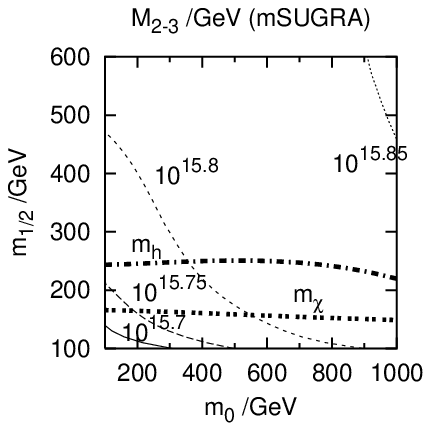}} & 
   \resizebox{6.5cm}{!}{\includegraphics{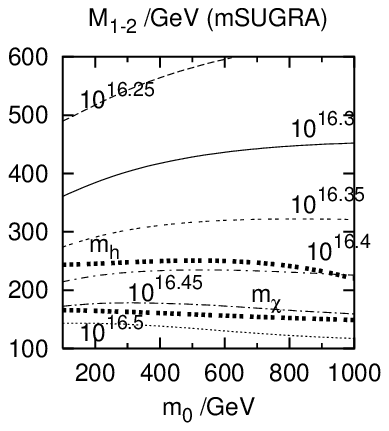}} \\
   \resizebox{7cm}{!}{\includegraphics{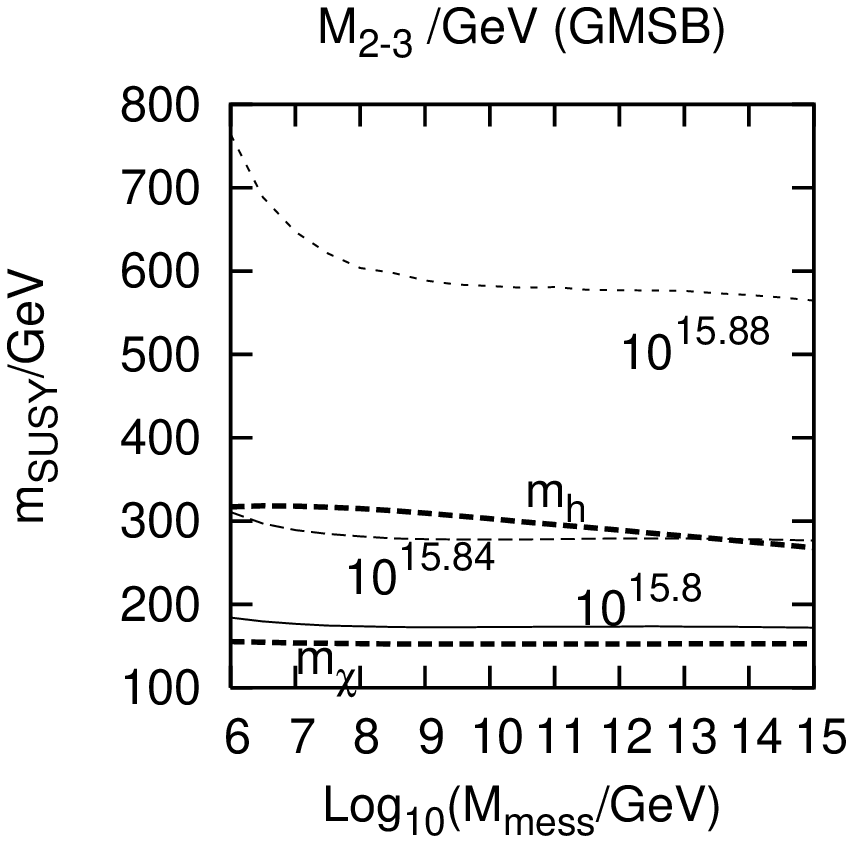}} &
   \resizebox{6.3cm}{!}{\includegraphics{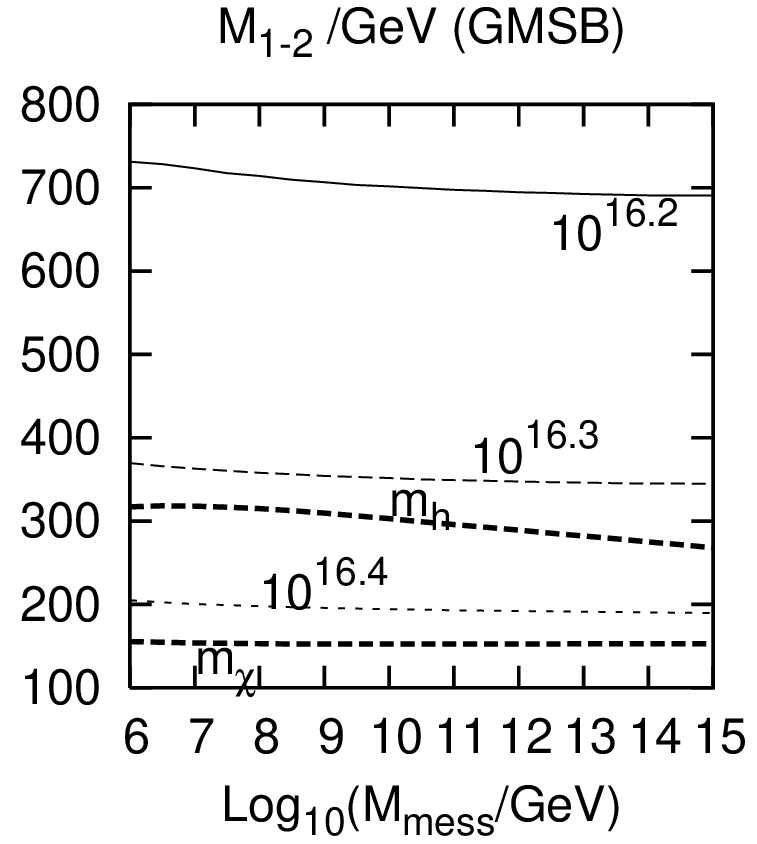}} 
  \end{tabular}
\caption{The left panels show the  contour plots of the energy scale 
$M_{2-3}$,  where the SU(2)$_L$ and SU(3)$_C$ gauge coupling constants 
become the same. 
The right panels show those of the energy scale $M_{1-2}$, where
the U(1)$_Y$ and SU(2)$_L$ gauge coupling constants become the same. 
The upper panels are contour plots on the $(m_0,m_{1/2})$ 
parameter space of mSUGRA SUSY breaking, 
the lower ones are for the 
$(M_{\rm mess},m_{\rm SUSY} \equiv ((1/24)/(4\pi))\Lambda)$ 
parameter space of the GMSB. 
Other parameters are fixed for both SUSY breakings;
$A_0 = 0$ GeV for mSUGRA SUSY breaking, and $\tan \beta = 10.0$ and
$\mu > 0$ for both SUSY breakings. 
$\alpha_s^{\overline{\rm MS},(5)}(M_Z)=0.1172$ is used 
as the QCD coupling constant in this figure. 
See the caption for Fig. 4 for more details 
about the region excluded by the LEP II experiments.   
}
\label{fig:mSUGRA-GMSB}
 \end{center}
\end{figure}

Figure \ref{fig:mSUGRA-GMSB} shows how $M_{2-3}$ and $M_{1-2}$ vary 
over the parameter space of the GMSB. 
The parameter space is spanned by two parameters: 
an overall mass scale $\Lambda$ of the SUSY breaking in the MSSM
sector and the messenger mass $M_{\rm mess}$. 
We assume that the messenger sector consists of one pair of SU(5)$_{\rm
GUT}$-($\bf 5$+$\bf 5^*$) representations.
Gaugino masses are given by
\begin{eqnarray}
 m_{\tilde{g}_i}
  &=& \frac{\alpha_i}{4\pi}\Lambda \left(1+{\cal O}
\left(\frac{\Lambda}{M_{\rm mess}}\right) \right) 
   \qquad (i=1,2,3) 
\end{eqnarray}
at the threshold $M_{\rm mess}$. 
We calculate the SUSY-particle spectrum, the SUSY threshold corrections to
the MSSM gauge coupling constants and the renormalization-group
evolution to the messenger scale using the code \cite{SOFTSUSY}. 
We include contributions from the messenger particles into the beta
functions in the renormalization-group evolution from the messenger
scale to the GUT scale.
$M_{2-3}$ and $M_{1-2}$ are obtained and are shown in Fig.
\ref{fig:mSUGRA-GMSB}. 
It is clear from Fig. \ref{fig:mSUGRA-GMSB} that 
the ranges of $M_{2-3}$ and $M_{1-2}$ are almost the same 
in the mSUGRA and GMSB parameter spaces.
Therefore, we conclude that there is little effect that comes purely
from the difference between the SUSY-particle spectra 
of the mSUGRA and GMSB. 

The gaugino masses satisfy the GUT relation in both the mSUGRA and GMSB
spectra, which may be the reason why $M_{2-3}$ and $M_{1-2}$ are almost
the same in the two spectra.
The gaugino mass spectrum, however, might not satisfy the GUT
relation\footnote{Gaugino masses without the GUT relation are
not unnatural at all in the product-group unification models 
we discuss in this article \cite{ACM}.}.
Even in this case, we can obtain the upper bound of the lifetime
through $M_{2-3}$ for the SU(5)$_{\rm GUT}\times$U(2)$_{\rm H}$ model
and through $M_{1-2}$ for the SU(5)$_{\rm GUT}\times$U(3)$_{\rm H}$
model. 

\subsection{Vector-Like SU(5)$_{\rm GUT}$-Multiplet at Low Energy}
\label{ssec:vect}

There are several motivations to consider charged particles 
in vector-like representations, whose masses are of the order 
of  
the SUSY-breaking scale or an intermediate scale.
Messenger particles are necessary in the GMSB models, 
and the anomaly cancellation of the discrete R symmetry 
also requires \cite{KMY} extra particles 
such as SU(5)$_{\rm GUT}$-({\bf 5}+{\bf 5}$^*$).

There are three effects on the proton lifetime 
in the presence of these particles. 
The first two effects come from the change in the values of the unified 
coupling constant $\alpha_{\rm GUT}$ and the renormalization factor 
$A_R$ of the proton-decay operators.
First, the unified gauge coupling constant is larger 
in the presence of new particles, and hence the decay rate is enhanced. 
Then, the lifetime is shortened by a factor not smaller than 0.66 
when a vector-like pair SU(5)$_{\rm GUT}$-({\bf 5}+{\bf 5}$^*$) 
exists at an energy scale not lower than 1 TeV. 
Second, the renormalization factor $A_R$ is changed by such a vector-like pair
only in its short-distance part. 
The new expression for $A^{(\rm SD)}_R$ is now given by 
\begin{eqnarray}
 A_R^{(\rm SD)} &=& 
\left(\frac{\alpha_C(M_Z)}{\alpha_C(M)}\right)^{\frac{4/3}{3}}
\left(\frac{\alpha_L(M_Z)}{\alpha_L(M)}\right)^{\frac{3/2}{-1}}
\left(\frac{\alpha_Y(M_Z)}{\alpha_Y(M)}\right)^{\frac{23/30}{-33/5}}
\nonumber\\
& &\times
\left(\frac{\alpha_C(M)}{\alpha_C(M_G)}\right)^{\frac{4/3}{2}}
\left(\frac{\alpha_L(M)}{\alpha_L(M_G)}\right)^{\frac{3/2}{-2}}
\left(\frac{\alpha_Y(M)}{\alpha_Y(M_G)}\right)^{\frac{23/30}{-38/5}}, 
\end{eqnarray}
where $M$ is the mass scale of the vector-like pair.
We find that $A_R^{(\rm SD)}$ increases from 2.1 to 2.5 as the mass
scale $M$ decreases from the GUT scale to 1 TeV.
Thus, the lifetime is shortened by a factor not smaller than 0.71 
because of the renormalization factor.

The third effect is due to threshold corrections from the vector-like
particles. The triplets and doublets in the vector-like pair 
{\bf 5}+{\bf 5}$^*$ are expected to have different masses, just as the 
bottom quark and tau lepton do. 
The triplets will be heavier than the doublets by 
\begin{equation}
 \frac{M_{{\bf 3}+{\bf 3}^*}}{M_{{\bf 2}+{\bf 2}^*}} \simeq 
  \left(\frac{\alpha_C(M_{{\bf 5}+{\bf 5}^*})}{\alpha_C(M_G)}\right)^{\frac{4}{3}},
\end{equation}
which increases from 1.0 to 2.1 as the mass scale $M_{{\bf 5+5^*}}$ of a
vector-like pair 
SU(5)$_{\rm GUT}$-({\bf 5}+{\bf 5}$^*$) decreases from the GUT scale 
to 1 TeV. 
The upper bound of the proton lifetime 
in the SU(5)$_{\rm GUT}\times$U(2)$_{\rm H}$ model becomes tighter by
a factor $(M_{{\bf 2}+{\bf 2}^*}/M_{{\bf 3}+{\bf 3}^*}) \gsim 0.48$ 
as $M_{2-3}$ is decreased by 
a factor $(M_{{\bf 2}+{\bf 2}^*}/M_{{\bf 3}+{\bf 3}^*})^{1/4}$;
in the SU(5)$_{\rm GUT}\times$U(3)$_{\rm H}$ model, instead, it 
is loosened by a factor 
$(M_{{\bf 3}+{\bf 3}^*}/M_{{\bf 2}+{\bf 2}^*})^{2/7} \lsim 1.2$ 
as $M_{1-2}$ is increased.

The proton decay is, thus, enhanced by all three effects in the
SU(5)$_{\rm GUT}\times$U(2)$_{\rm H}$ model; the lifetime is shortened 
by a factor of 0.22 when SU(5)$_{\rm GUT}$-({\bf 5}+{\bf 5}$^*$) exists 
at 1 TeV.
The rate is enhanced also 
in the SU(5)$_{\rm GUT}\times$U(3)$_{\rm H}$ model; 
the lifetime is shortened by a factor of 0.56.


\section{Conclusions and Discussion}

\label{sec:conclusion}



We analysed the proton-decay  amplitude in a class of models of SUSY
GUTs:
SU(5)$_{\rm GUT}$ $\times$U($N$)$_{\rm H}$ models with $N=2,3$.
Dimension-five proton-decay operators are completely forbidden, 
and hence the gauge-boson exchange is the process 
that dominates the proton decay. 
We found that the gauge-boson mass is bounded from above by 
\begin{equation}
M_G \lsim 0.40 \times 10^{-0.82 \times c} \times M_{2-3}
\end{equation}
in the SU(5)$_{\rm GUT}\times$U(2)$_{\rm H}$ model\footnote{
This expression for the upper bound of $M_G$ is valid as long as
$c\gsim -0.16+1.3 \times(\epsilon_g/\alpha_{\rm GUT})$.} and by 
\begin{equation}
M_G \lsim 0.60 \times 10^{-0.4 \times c} \times M_{1-2}
\end{equation}
in the SU(5)$_{\rm GUT}\times$U(3)$_{\rm H}$ model.
Here, $M_{2-3}$ ($M_{1-2}$) denotes an energy scale where SU(2)$_L$ and
SU(3)$_C$  
(U(1)$_Y$ and SU(2)$_L$) gauge coupling constants are equal, respectively.
In the right-hand sides, $c$ are coefficients 
of non-renormalizable operator (\ref{eq:non-ren}) 
in the SU(5)$_{\rm GUT} \times$U(2)$_{\rm H}$ model and of the one 
that corresponds to (\ref{eq:non-ren}) 
in the SU(5)$_{\rm GUT} \times$U(3)$_{\rm H}$ model.
It is quite important to note that the upper bound was obtained in these
models (for fixed $c$), which leads to the upper bound of the lifetime.
Although the gauge-boson masses are bounded also from below in the latter
model, the lower bound  is of no importance. 
This is because it predicts a lifetime much shorter 
than the lower bound obtained so far from experiments. 

The coefficients $c$ directly affect the gauge coupling unification, 
and hence they appear in the above formulae.
One will see later that they are the largest source of uncertainties 
in the upper bound of the lifetime if $c$ are of the order of unity.
Although there may be an extra (broken) symmetry or any dynamics that
suppress the non-renormalizable operators, we leave $c$ in the formulae 
for generic cases.   

In section \ref{sec:intro}, we briefly mentioned two other classes of
 models of SUSY GUTs constructed in four-dimensional space-time.
Let us make a brief summary on the mass of the gauge bosons 
of such models before we proceed to a discussion of the lifetime.

Let us first discuss the gauge-boson mass in the models 
in \cite{BDS,Barr}.
The spectrum around the GUT scale consists of 
three (({\bf adj.},{\bf 1})$^0$ + ({\bf 1},{\bf adj})$^0$)'s 
and two (({\bf 3},{\bf 2})$^{-5/6}$ + h.c.)'s of the MSSM gauge group, in
addition to the GUT gauge boson.   
Parameters of the models allow a spectrum where the matter particles are
lighter than the GUT gauge boson.
Then, 1-loop threshold corrections from such a spectrum imply that the
GUT-gauge-boson mass is heavier than the energy scale of  approximate
unification \cite{nodw}.    
Therefore, no upper bound is virtually obtained in the models in
\cite{BDS,Barr}. 
A lower bound might be obtained, yet no full study has been done so far.
Non-renormalizable operators in the gauge kinetic functions affect the
matching equations just as in our analysis. 

On the contrary, in the models in \cite{Maekawa}, non-renormalizable
operators do not affect the matching equations and, moreover, the mass
of the GUT gauge boson is smaller than the energy scale where the three
gauge coupling  constants are approximately unified: 
\begin{equation}
M_G \sim \lambda^a M_{\rm unif}, 
\end{equation}
where $M_{\rm unif} \sim M_{2-3} \sim M_{1-2}$, $\lambda$ a small
parameter of the order of $10^{-1}$ and $a$ the charge of a field whose
VEV breaks the SU(5)$_{\rm GUT}$ symmetry down to
SU(3)$_C\times$SU(2)$_L\times$U(1)$_Y$. 
Thus, $M_G$ is fairly small in the models.
The upper bound would be obtained once a model ($\lambda$ and $a$, in
particular) is fixed. 
The Super-Kamiokande experiment already puts constraints on the choice of
$\lambda$ and $a$. 
The proton decays also through dimension-five operators, 
although these operators can be suppressed in some models in this class.

Thus, the ranges of the proton lifetime of those models lie 
in the following order:   
\begin{equation}
\tau(\cite{Maekawa})_{\rm dim.6} \sim 
\tau(\SU(5)_{\rm GUT}\times\U(2)_{\rm H}) \lsim 
\tau(\SU(5)_{\rm GUT}\times\U(3)_{\rm H}) \lsim 
\tau(\cite{BDS},\cite{Barr}). 
\end{equation}
However, the ranges would have certain amount of overlap 
between one another, 
and hence it would be impossible to single out a model only from 
the decay rate of the proton.  
Detailed information on the branching ratio of various decay modes 
does not help for that purpose either; the decay is induced 
in all the above models\footnote{
The proton decay can be induced by the gauge-boson exchange 
also in SUSY-GUT models in higher-dimensional spacetime 
\cite{Kawamura,WittenG2}.
The branching ratio of various modes can be different \cite{HN2,FrW} 
from the standard one in those models.} 
by one and the same\footnote{If the
dimension-five decay is not the dominant process 
in the last class of models.} mechanism: the gauge-boson exchange.

Even if no model can be singled out, one can, and 
one will be able to exclude some of the models on the basis of 
experimental results 
currently available and obtained in the future, respectively.
We summarize, in the following, the upper bound of the proton lifetime 
for the SU(5)$_{\rm GUT}\times$U(2)$_{\rm H}$ model 
and the SU(5)$_{\rm GUT}\times$U(3)$_{\rm H}$ model.  
It would also be of importance if one finds an upper bound and a lower
bound of the lifetime in models in \cite{Maekawa} and in
\cite{BDS,Barr,DNS}, respectively. 

Now the proton lifetime is bounded from above by
\begin{equation}
\tau(p \rightarrow \pi^0 e^+) \lsim 4.1 \times 10^{32} 
    \left(\frac{M_{2-3}} {10^{15.8} {\rm GeV}} \right)^4 
   \left(10^{-3.2 c}\right)
   \left( \frac{0.015 {\rm GeV}^3}{\alpha_{\rm H}} \right)^2
   \left( \frac{ 1 }{25\alpha_{\rm GUT}} \right)^2 
   \left( \frac{2.5}{A_R} \right)^2 {\rm yrs}
\end{equation}
in the SU(5)$_{\rm GUT} \times$U(2)$_{\rm H}$ model.
The largest uncertainty in this prediction comes from the value of $c$.
Another from the systematic error in $\alpha_{\rm H}$. 
No estimate is available for the value of $c$. 
The error in $\alpha_{\rm H}$ is not studied very much, 
yet the lifetime is changed by a factor of (0.5--2) 
if the conservative estimate in \cite{Raby} is adopted\footnote{
We presented a numerical value of the upper bound of the proton lifetime 
in the abstract of this article. 
These two uncertainties are not included there, since 
it is impossible to make a precise estimate of them at this moment.
The following two uncertainties, on the other hand, are included 
in obtaining the numerical value in the abstract.}.
The experimental value of the QCD coupling constant 
$\alpha_s^{\overline{\rm MS},(5)}(M_Z)$ changes the prediction 
through the change in $M_{2-3}$.
The upper bound is changed by a factor of (0.15--5.9) 
when the coupling constant is varied by $\pm 2 \sigma$ error 
determined by experiments.  
The threshold corrections from SUSY particles also changes the
prediction through the change in $M_{2-3}$.
They change $M_{2-3}$ typically from $10^{15.77}$ GeV 
to $10^{15.90}$ GeV,
and hence the upper bound is changed by a factor of (0.75--2.5).  
Therefore, the theoretical upper bound exceeds the experimental 
lower bound ($\tau(p \rightarrow \pi^0 e^+) \gsim 5 \times 10^{33}$ yrs;
90\% C.L.) only when\footnote{This statement holds 
as long as the non-renormalizable operator (\ref{eq:non-ren}) 
is neglected, i.e. as long as $c \gg -1$. 
} all the low-energy uncertainties are exploited. 
See section \ref{ssec:uncert2} for the uncertainties 
that arise in the way of our analysis. 
 
The lifetime is shortened by a factor not smaller than 0.22
if SU(5)$_{\rm GUT}$-({\bf 5}+{\bf 5}$^*$) exists at low energy.
The threshold corrections from these particles contribute 
by a factor not smaller than 0.47 through the change in $M_{2-3}$, 
and the changes in $\alpha_{\rm GUT}$ and in $A_R$ contribute 
by factors not smaller than 0.66 and 0.71, respectively.  
Thus, those particles at 1 TeV would hardly be reconciled 
with the experimetal bound without incorporating the non-renormalizable
operator (\ref{eq:non-ren}). 

The lifetime is bounded from above by 
\begin{equation}
\tau( p \rightarrow \pi^0 e^+) \lsim 2.1  \times 10^{35} 
\left( \frac{M_{1-2}}{10^{16.3} \GEV} \right)^4
\left(10^{- 2 c}\right)
\left( \frac{0.015 \GEV^3}{\alpha_{\rm H}} \right)^2
\left( \frac{1}{25 \alpha_{\rm GUT}} \right)^2
\left( \frac{2.5}{A_R} \right)^2  {\rm yrs}
\end{equation}
in the SU(5)$_{\rm GUT}\times$U(3)$_{\rm H}$ model.
Uncertainties arise\footnote{They are not included in obtaining the
numerical value in the abstract just for the same reason as in the
previous model.} from $c$ and $\alpha_{\rm H}$ as in the previous model.  
The value of the QCD coupling constant is not relevant to the prediction.
The SUSY threshold corrections changes the upper bound typically 
by a factor of (0.40--2.5) as $M_{1-2}$ changes 
from $10^{16.20}$ GeV to $10^{16.40}$ GeV. 

The lifetime is shortened by a factor not smaller than 0.56 
in the presence of SU(5)$_{\rm GUT}$-({\bf 5}+{\bf 5}$^*$) 
below the GUT scale. 
The decay is more enhanced as their mass is smaller. 
The enhancement factor 0.56 (when the mass is 1 TeV) consists of 
suppression factor 1.2, which comes from $M_{1-2}$ changed by threshold
corrections of these particles, and enhancement factors 0.66 and 0.71 
respectively from $\alpha_{\rm GUT}$ and $A_R$.

\section*{Acknowledgements}
The authors are grateful to the Theory Division of CERN 
for the hospitality, where earlier part of this work was done. 
They thank T. Yanagida for discussions and a careful reading of this
manuscript.  
T.W. thanks the Japan Society for the Promotion of Science for financial
support.

\appendix

\section{Renormalization-Group Equations}
\label{sec:RGE}

In this section, renormalization-group equations of coupling constants of the
models are listed.

{\bf SU(5)$_{\rm GUT}\times$U(2)$_{\rm H}$ model}

\begin{eqnarray}
 \frac{\partial}{\partial \ln \mu}\left( \frac{1}{\alpha_{\rm 2H}}(\mu)\right)
& = & \frac{-2}{2\pi} \qquad \qquad \qquad \mbox{(1-loop)}  
        \label{eq:RGE-U2-2H} \\
& - & \frac{2}{2\pi}\frac{(6\alpha_{\rm 2H}-5\alpha_{\rm 2H}^\lambda
                           -\alpha_{\rm 2H}^{\lambda'})}{2\pi} 
                 \nonumber \\
& - & \frac{1}{2\pi}\frac{6(3\alpha_{\rm 2H}+\alpha_{\rm 1H})
               -5 (3\alpha_{\rm 2H}^\lambda+\alpha_{\rm 1H}^\lambda)
               - (3\alpha_{\rm 2H}^{\lambda'}+\alpha_{\rm 1H}^{\lambda'})}
                          {4\pi} 
                 \nonumber \\
&  & - \frac{5}{2\pi}\left(\frac{\frac{24}{10}}{\pi}\alpha_{\rm GUT}\right). 
        \nonumber  \\  
 \frac{\partial}{\partial \ln \mu}
  \left( \frac{1}{\alpha_{\rm 2H}^\lambda}(\mu)\right)
& = & \frac{-2}{2\pi} \left(\frac{\alpha_{\rm 2H}}
                                 {\alpha_{\rm 2H}^\lambda}\right)
 + \frac{1}{\alpha_{\rm 2H}^\lambda}
                 \frac{(6\alpha_{\rm 2H}-5\alpha_{\rm 2H}^\lambda
                      -\alpha_{\rm 2H}^{\lambda'})}{2\pi} 
      \label{eq:RGE-U2-2HL}\\
& + & \frac{2}{\alpha_{\rm 2H}^\lambda}
           \frac{(3\alpha_{\rm 2H}+\alpha_{\rm 1H})
                -(3\alpha_{\rm 2H}^\lambda+\alpha_{\rm 1H}^\lambda)}{4\pi} 
                 \nonumber \\
&& + \frac{2}{\alpha_{\rm 2H}^\lambda}
       \left(\frac{\frac{24}{10}}{\pi}\alpha_{\rm GUT} \right).   \nonumber \\
 \frac{\partial}{\partial \ln \mu}
  \left( \frac{1}{\alpha_{\rm 2H}^{\lambda'}}(\mu)\right) 
& = &  \frac{-2}{2\pi} \left(\frac{\alpha_{\rm 2H}}
                                 {\alpha_{\rm 2H}^{\lambda'}}\right)
 + \frac{1}{\alpha_{\rm 2H}^{\lambda'}}
                 \frac{(6\alpha_{\rm 2H}-5\alpha_{\rm 2H}^\lambda
                      -\alpha_{\rm 2H}^{\lambda'})}{2\pi}  
        \label{eq:RGE-U2-2HLp} \\
& + & \frac{2}{\alpha_{\rm 2H}^{\lambda'}}
           \frac{(3\alpha_{\rm 2H}+\alpha_{\rm 1H})
             -(3\alpha_{\rm 2H}^{\lambda'}+\alpha_{\rm 1H}^{\lambda'})}{4\pi}. 
                 \nonumber \\ 
 \frac{\partial}{\partial \ln \mu}\left( \frac{1}{\alpha_{\rm 1H}}(\mu)\right)
& = &  \frac{-6}{2\pi} \qquad \qquad \qquad \mbox{(1-loop)}  
         \label{eq:RGE-U2-1H} \\
& - &  \frac{1}{2\pi}\frac{6(3\alpha_{\rm 2H}+\alpha_{\rm 1H})
               -5 (3\alpha_{\rm 2H}^\lambda+\alpha_{\rm 1H}^\lambda)
               - (3\alpha_{\rm 2H}^{\lambda'}+\alpha_{\rm 1H}^{\lambda'})}
                          {4\pi} 
                 \nonumber \\
&  & - \frac{5}{2\pi}\left(\frac{\frac{24}{10}}{\pi}\alpha_{\rm GUT}\right). 
        \nonumber   \\
 \frac{\partial}{\partial \ln \mu}
  \left( \frac{1}{\alpha_{\rm 1H}^\lambda}(\mu)\right) 
& = &  \frac{-6}{2\pi} \left(\frac{\alpha_{\rm 1H}}
                                 {\alpha_{\rm 1H}^\lambda}\right)
 + \frac{1}{\alpha_{\rm 1H}^\lambda}
                 \frac{(6\alpha_{\rm 1H}-5\alpha_{\rm 1H}^\lambda
                      -\alpha_{\rm 1H}^{\lambda'})}{2\pi} 
           \label{eq:RGE-U2-1HL} \\
& + & \frac{2}{\alpha_{\rm 1H}^\lambda}
           \frac{(3\alpha_{\rm 2H}+\alpha_{\rm 1H})
                -(3\alpha_{\rm 2H}^\lambda+\alpha_{\rm 1H}^\lambda)}{4\pi} 
                 \nonumber \\
&& + \frac{2}{\alpha_{\rm 1H}^\lambda}
       \left(\frac{\frac{24}{10}}{\pi}\alpha_{\rm GUT} \right).   \nonumber \\
 \frac{\partial}{\partial \ln \mu}
  \left( \frac{1}{\alpha_{\rm 1H}^{\lambda'}}(\mu)\right) 
& = &  \frac{-6}{2\pi} \left(\frac{\alpha_{\rm 1H}}
                                 {\alpha_{\rm 1H}^{\lambda'}}\right)
 + \frac{1}{\alpha_{\rm 1H}^{\lambda'}}
                 \frac{(6\alpha_{\rm 1H}-5\alpha_{\rm 1H}^\lambda
                      -\alpha_{\rm 1H}^{\lambda'})}{2\pi} 
         \label{eq:RGE-U2-1HLp} \\
& + & \frac{2}{\alpha_{\rm 1H}^{\lambda'}}
           \frac{(3\alpha_{\rm 2H}+\alpha_{\rm 1H})
             -(3\alpha_{\rm 2H}^{\lambda'}+\alpha_{\rm 1H}^{\lambda'})}{4\pi}. 
                 \nonumber 
\end{eqnarray}

{\bf SU(5)$_{\rm GUT}\times$U(3)$_{\rm H}$ model}

\begin{eqnarray}
 \frac{\partial}{\partial \ln \mu}\left( \frac{1}{\alpha_{\rm 3H}}(\mu)\right)
    &=& 0  \qquad \qquad \qquad \mbox{(1-loop)} \\
& - & \frac{3}{2\pi}\frac{(6\alpha_{\rm 3H}-5\alpha_{\rm 3H}^\lambda
                     -\alpha_{\rm 3H}^{\lambda'})}{2\pi}    \nonumber \\
& - & \frac{1}{2\pi}\frac{6(8\alpha_{\rm 3H}+\alpha_{\rm 1H})
               -5(8\alpha_{\rm 3H}^\lambda+\alpha_{\rm 1H}^\lambda)
               -(8\alpha_{\rm 3H}^{\lambda'}+\alpha_{\rm 1H}^{\lambda'})}
                        {6\pi}   \nonumber \\
&& + \left(-\frac{5}{2\pi}\frac{\frac{24}{10}}{\pi}\alpha_{\rm GUT}
           +\frac{10}{2\pi}\frac{1}{2\pi}\alpha_h \right). \nonumber \\
 \frac{\partial}{\partial \ln \mu}
  \left( \frac{1}{\alpha_{\rm 3H}^\lambda}(\mu)\right)  
& = & 0 + \frac{1}{\alpha_{\rm 3H}^\lambda}
                 \frac{(6\alpha_{\rm 3H}-5\alpha_{\rm 3H}^\lambda 
                                         -\alpha_{\rm 3H}^{\lambda'})}{2\pi} \\
& + & \frac{2}{\alpha_{\rm 3H}^\lambda}
           \frac{(8\alpha_{\rm 3H}+\alpha_{\rm 1H})
                -(8\alpha_{\rm 3H}^\lambda+\alpha_{\rm 1H}^\lambda)}{6\pi} 
                 \nonumber \\
&& + \frac{2}{\alpha_{\rm 3H}^\lambda}
       \left(\frac{\frac{24}{10}}{\pi}\alpha_{\rm GUT}
            -\frac{1}{2\pi}\alpha_h
       \right).               \nonumber \\
 \frac{\partial}{\partial \ln \mu}
  \left( \frac{1}{\alpha_{\rm 3H}^{\lambda'}}(\mu)\right)  
& = & 0 + \frac{1}{\alpha_{\rm 3H}^{\lambda'}}
                 \frac{(6\alpha_{\rm 3H}-5\alpha_{\rm 3H}^\lambda
                                         -\alpha_{\rm 3H}^{\lambda'})}{2\pi} \\
& + & \frac{2}{\alpha_{\rm 3H}^{\lambda'}}
        \frac{(8\alpha_{\rm 3H}+\alpha_{\rm 1H})
             -(8\alpha_{\rm 3H}^{\lambda'}+\alpha_{\rm 1H}^{\lambda'})}{6\pi} 
                 \nonumber \\
&& + \frac{2}{\alpha_{\rm 3H}^{\lambda'}}
       \left(\qquad \qquad \qquad 
            -\frac{5}{2\pi}\alpha_h
       \right).               \nonumber \\
 \frac{\partial}{\partial \ln \mu}\left( \frac{1}{\alpha_{\rm 1H}}(\mu)\right)
& = & \frac{-6}{2\pi} \qquad \qquad \qquad \mbox{(1-loop)} \\
& - & \frac{1}{2\pi}\frac{6(8\alpha_{\rm 3H}+\alpha_{\rm 1H})
               -5(8\alpha_{\rm 3H}^\lambda+\alpha_{\rm 1H}^\lambda)
               -(8\alpha_{\rm 3H}^{\lambda'}+\alpha_{\rm 1H}^{\lambda'})}
                         {6\pi}   \nonumber \\
&& + \left(-\frac{5}{2\pi}\frac{\frac{24}{10}}{\pi}\alpha_{\rm GUT}
           +\frac{10}{2\pi}\frac{1}{2\pi}\alpha_h \right). \nonumber   \\
 \frac{\partial}{\partial \ln \mu}
  \left( \frac{1}{\alpha_{\rm 1H}^\lambda}(\mu) \right) 
& = & \frac{-6}{2\pi}\left(\frac{\alpha_{\rm 1H}}{\alpha_{\rm 1H}^\lambda} 
                     \right)  
+ \frac{1}{\alpha_{\rm 1H}^\lambda}
                 \frac{(6\alpha_{\rm 1H}-5\alpha_{\rm 1H}^\lambda 
                                         -\alpha_{\rm 1H}^{\lambda'})}{2\pi} \\
& + & \frac{2}{\alpha_{\rm 1H}^\lambda}
           \frac{(8\alpha_{\rm 3H}+\alpha_{\rm 1H})
                -(8\alpha_{\rm 3H}^\lambda+\alpha_{\rm 1H}^\lambda)}{6\pi} 
                 \nonumber \\
&& + \frac{2}{\alpha_{\rm 1H}^\lambda}
       \left(\frac{\frac{24}{10}}{\pi}\alpha_{\rm GUT}
            -\frac{1}{2\pi}\alpha_h
       \right).               \nonumber \\
 \frac{\partial}{\partial \ln \mu}
  \left( \frac{1}{\alpha_{\rm 1H}^{\lambda'}}(\mu)\right) 
& = & \frac{-6}{2\pi}\left(\frac{\alpha_{\rm 1H}}{\alpha_{\rm 1H}^{\lambda'}} 
                     \right) 
+ \frac{1}{\alpha_{\rm 1H}^{\lambda'}}
                 \frac{(6\alpha_{\rm 1H}-5\alpha_{\rm 1H}^\lambda 
                                         -\alpha_{\rm 1H}^{\lambda'})}{2\pi} \\
& + & \frac{2}{\alpha_{\rm 1H}^{\lambda'}}
         \frac{(8\alpha_{\rm 3H}+\alpha_{\rm 1H})
              -(8\alpha_{\rm 3H}^{\lambda'}+\alpha_{\rm 1H}^{\lambda'})}{6\pi} 
                 \nonumber \\
&& + \frac{2}{\alpha_{\rm 1H}^{\lambda'}}
       \left(\qquad \qquad \qquad 
            -\frac{5}{2\pi}\alpha_h
       \right).               \nonumber  \\
  \frac{\partial}{\partial \ln \mu} \left( \frac{1}{\alpha_{h}}(\mu)\right) 
& = & \frac{1}{\alpha_h}
           \frac{2(8\alpha_{\rm 3H}+\alpha_{\rm 1H})
                -(8\alpha_{\rm 3H}^\lambda+\alpha_{\rm 1H}^\lambda)
                -(8\alpha_{\rm 3H}^{\lambda'}+\alpha_{\rm 1H}^{\lambda'})}
                {6\pi}  \\
&& + \left(\frac{1}{\alpha_h}\right)
         \left(\frac{2 \times \frac{24}{10}}{\pi}\alpha_{\rm GUT}
              -\frac{(1+5+3)}{2\pi}\alpha_h\right). \nonumber
\end{eqnarray}

\section{${\cal N}$ = 2 SUSY and Infrared-Fixed \\Renormalization-Group Flow}
\label{sec:U2-N2}

\begin{figure}[ht]
\begin{center}
\begin{minipage}{0.48\linewidth}
\begin{center}
 \includegraphics[width=\linewidth,height=0.8\linewidth]{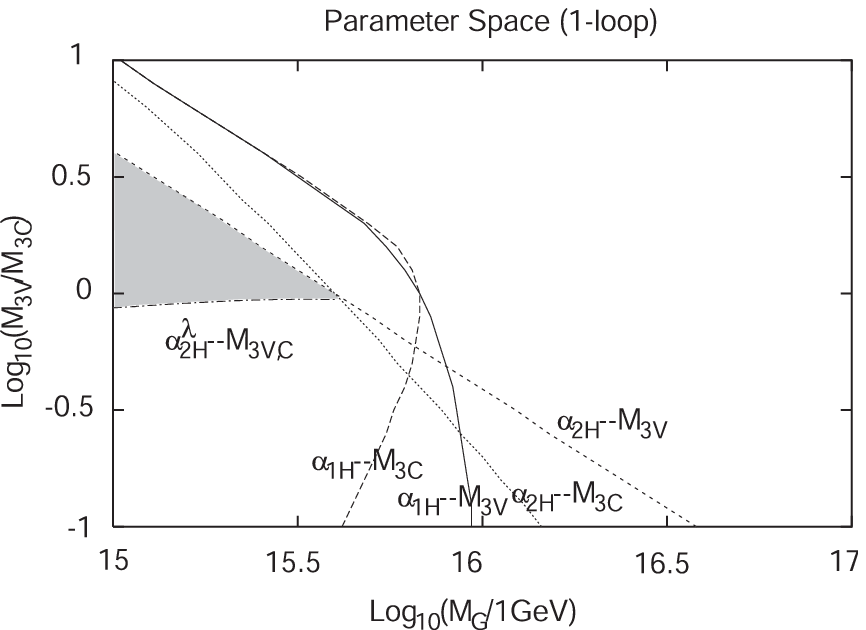} 
\end{center}
\end{minipage}
\begin{minipage}{0.48\linewidth}
\begin{center}
 \includegraphics[width=\linewidth,height=0.8\linewidth]{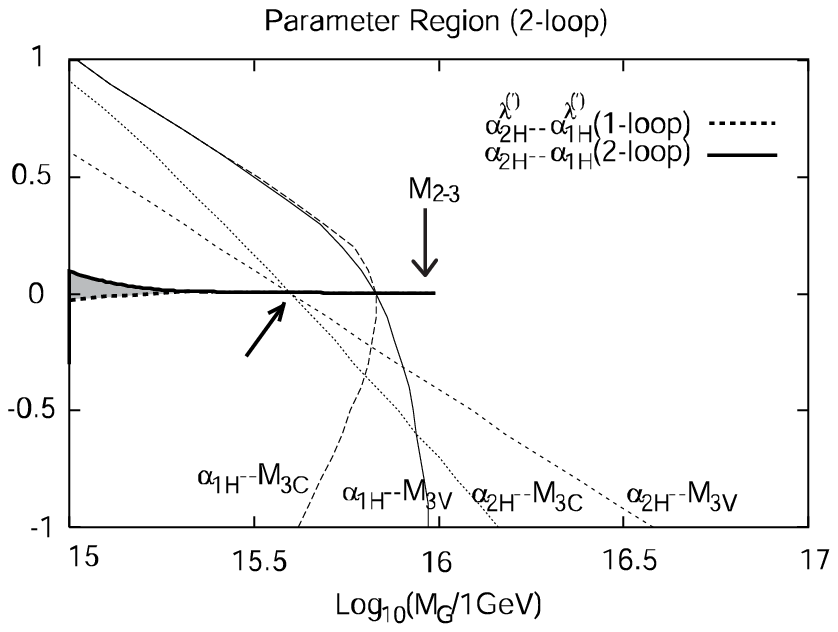} 
\end{center}
\end{minipage}
\caption{Parameter region of the SU(5)$_{\rm GUT}\times$U(2)$_{\rm H}$
model. 
The parameter space of the model spanned by two free parameters 
$M_G$ and $M_{3V}/M_{3C}$ are restricted by requiring that all the 
running coupling constants of the model remain finite
while the renormalization point is below the  heaviest particle 
of the model.
The left panel shows the parameter region, where the 1-loop
 renormalization group is used for all the coupling constants.
The right-hand sides of the four curves labelled 
 ``(gauge-coupling)-mass'' are excluded.
The region below a curve labelled  
``$\alpha_{\rm 2H}^\lambda$--$M_{3V,C}$'' is also excluded. 
Thus, the parameter space of the model is restricted to the shaded
 triangular region.
The right panel shows the parameter region (shaded), 
where 2-loop effects are included in the renormalization-group
 equations of gauge coupling constants.  
The four curves are those found in the left panel;
we keep them just because they make it easier to compare the panel with
 the left one. 
The majority of the triangular region in the left panel  
is further excluded because of the 2-loop effects, 
and only a small region survives near the line $M_{3V}\simeq M_{3C}$.
The upper bound of $M_G$ is indicated by an arrow.
In the right panel, $M_{2-3}$ indicates the unification point between
 $1/\alpha_{2}$ and $1/\alpha_3(+2\sigma)$ (see Fig. \ref{fig:closeup}
 for details). 
It is easy to see that $M_G \lsim 10^{15.6}$ GeV 
$\simeq$ ($10^{-0.4} \simeq 0.40$) $\times$ 
($M_{2-3} \simeq 10^{16.0}$ GeV). 
Both two panels use $\alpha_s^{\overline{\rm MS},(5)}(M_Z) = 0.1212$.
The effects from the non-renormalizable operator (\ref{eq:non-ren}) 
are not included here.  
}
\label{fig:U2Region}
\end{center}
\end{figure}

Particle contents in the GUT-symmetry-breaking sector of the SU(5)$_{\rm
GUT}\times$U(2)$_{\rm H}$ model can be regarded as multiplets of the
${\cal N}$ = 2 SUSY \cite{HY}, and ${\cal N}$ = 2 SUSY is enhanced in
this sector \cite{IWY} when the SU(5)$_{\rm GUT}$ gauge interaction is
neglected and coupling constants satisfy 
\begin{equation}
 g_{\rm 1H} \simeq  \lambda_{\rm 1H}( \sim \lambda_{\rm 1H}'), \qquad \qquad
 g_{\rm 2H} \simeq  \lambda_{\rm 2H}( \sim \lambda_{\rm 2H}').
\label{eq:N=2-relation}
\end{equation}
One can see in the right panel of Fig. \ref{fig:U2Region} that
the parameter region survives in the presence of the 2-loop effects only
when the ${\cal N}$ = 2 SUSY is approximately preserved; the $M_{3V}
\simeq M_{3C}$ line is equivalent to $\alpha_{\rm 2H} \simeq 
\alpha_{\rm 2H}^{\lambda^{(')}}$ when the SU(5)$_{\rm GUT}$ gauge
interaction is neglected. 

This is not a coincidence.
In any gauge theory with ${\cal N}$ = 2 SUSY, gauge coupling constants
are renormalized only at the 1-loop level \cite{N2-Gauge1lp}.
Anomalous dimensions of hypermultiplets vanish \cite{N2-Hyptree} at all
orders in perturbative expansion, and even non-perturbatively.
Therefore, the parameter allowed in the 1-loop analysis is still allowed
when ${\cal N}$ = 2 SUSY is preserved even after the 2-loop effects have 
also been taken into account.

The band of the region around the (${\cal N}$ = 2)-SUSY line almost becomes a
line as $M_G$ becomes larger.
For parameters above that line, the $\alpha_{\rm 2H}$ coupling constant
becomes large at a renormalization point lower than $M_{3V}$, while for 
parameters below the line, the $\alpha_{\rm 2H}^{\lambda}$ coupling
becomes large at a renormalization point lower than $M_{3C}$; 
a viable set of parameters was not found for large $M_G$ even {\em on} the line  
in our numerical calculation.
It does not mean, however, that the parameter does not exist at all 
on the (${\cal N}$ = 2)-SUSY line ($M_{3V} \simeq M_{3C}$) for large
$M_G$, as seen below. 
The (${\cal N}$ = 2)-SUSY relations (\ref{eq:N=2-relation}) are not only 
renormalization-group invariant but also infrared (IR)-fixed relations of
the renormalization group: 
\begin{eqnarray}
 \frac{\partial}{\partial \ln \mu}
   \left(\frac{1}{\alpha_{\rm 2H}}-\frac{1}{\alpha_{\rm 2H}^\lambda}\right)(\mu)
  & = & \left(\frac{21}{2\pi}+\frac{7}{\alpha_{\rm 2H}^\lambda}\right)
     \frac{(\alpha_{\rm 2H}^\lambda-\alpha_{\rm 2H})}{2\pi}
   + \left(\frac{6}{2\pi}+\frac{2}{\alpha_{\rm 2H}^\lambda}\right)
     \frac{(\alpha_{\rm 1H}^\lambda-\alpha_{\rm 1H})}{4\pi}, 
   \label{eq:U2-IR-RGE-2H}\\
 \frac{\partial}{\partial \ln \mu}
   \left(\frac{1}{\alpha_{\rm 1H}}-\frac{1}{\alpha_{\rm 1H}^\lambda}\right)(\mu)
  & = &  \left(\frac{6}{2\pi}+\frac{2}{\alpha_{\rm 1H}^\lambda}\right)
     \frac{3(\alpha_{\rm 2H}^\lambda-\alpha_{\rm 2H})
           +(\alpha_{\rm 1H}^\lambda-\alpha_{\rm 1H})}{4\pi}, 
    \label{eq:U2-IR-RGE-1H}
\end{eqnarray} 
where the SU(5)$_{\rm GUT}$ interaction is still neglected.
This implies that the renormalization-group evolution
to UV immediately becomes unstable\footnote{ 
This is the reason why we believe that it would not make the parameter
region wider even if we set the values of undetermined
parameters $\alpha_{\rm 2H}^{\lambda'}$ and 
$\alpha_{\rm 1H}^{\lambda^{(')}}$ differently from those 
in Eq. (\ref{eq:U2-initial}). 
A deviation from the (${\cal N}$ = 2)-SUSY relation at $M=M_G$ would
immediately lead to a UV-unstable behaviour in the renormalization-group
evolution.}, even for a set of parameters that is slightly distant from  
the IR-fixed relations.
The IR-fixed property (UV instability) also implies that the parameter
region is thin only when $M = M_G$, 
and not when the coupling constants are evaluated at $M \gg M_G$. 

Thus, we can expect that the 1-loop analysis is completely reliable for
a set of parameters {\em exactly on} the (${\cal N} = 2$)-SUSY
line and, in particular, that a viable set of parameters 
{\em does} exist on the
line even if it is not found in the numerical analysis. 
Therefore, the maximum value of $M_G$ is given at a point
indicated by an arrow in the right panel of Fig.
\ref{fig:U2Region}.  
At least, there would be no doubt that the maximum value of $M_G$
obtained in such a way provides a conservative upper bound of $M_G$.

The above argument, however, is correct 
only when the SU(5)$_{\rm GUT}$ gauge interaction is neglected.
Therefore, let us now discuss the effects of the SU(5)$_{\rm GUT}$ 
gauge interaction.
These break the ${\cal N}$ = 2 SUSY in the sector.
Thus, the (${\cal N}$ = 2)-SUSY relations in Eq. (\ref{eq:N=2-relation})
are no longer renormalization-group-invariant, and the
renormalization-group flow is no longer 1-loop exact.  
However, the SU(5)$_{\rm GUT}$ interaction is much weaker than the 
U(2)$_{\rm H}$ interactions, and its effects 
are small\footnote{
This can be seen from the fact that the parameter region 
is still almost around the (${\cal N}$ = 2)-SUSY line, i.e. 
$M_{3V}\simeq M_{3C}$, in the right panel of Fig. \ref{fig:U2Region}.
(${\cal N}$ = 2)-SUSY-breaking interactions are included in the figure.
}.
Thus, they can be treated as small perturbations to the (${\cal N}$ =
2)-SUSY flow. 
In particular, the IR-fixed property of the renormalization-group 
equations (\ref{eq:RGE-U2-2H})--(\ref{eq:RGE-U2-1HLp}) 
is not changed\footnote{
There is no IR-fixed relation in its strict meaning in the presence of
the  SU(5)$_{\rm GUT}$ interaction. 
The ``IR-fixed relations'' in Eq. (\ref{eq:IRfixed-wSU(5)}) involve  
$\alpha_{\rm 2H}^{\lambda^{(')}}$ and $\alpha_{\rm 1H}^{\lambda^{(')}}$
in the right-hand sides, and hence the ``fixed relations'' 
themselves change as the coupling constants flow. 
However, we still consider that they are almost IR-fixed relations,
because the beta functions of the quantities in the right-hand sides
($\simeq {\cal O}(\alpha_{\rm GUT})$) are much smaller than those of
the quantities in the left-hand sides.}, 
except that the IR-fixed relations are slightly modified into
\begin{equation}
 (\alpha_{\rm 2H}-\alpha_{\rm 2H}^\lambda), \;\;
 (\alpha_{\rm 2H}-\alpha_{\rm 2H}^{\lambda'}), \;\; 
 (\alpha_{\rm 1H}-\alpha_{\rm 1H}^\lambda), \;\; 
 (\alpha_{\rm 1H}-\alpha_{\rm 1H}^{\lambda'}) 
   \simeq {\cal O}(\alpha_{\rm GUT}). 
\label{eq:IRfixed-wSU(5)}
\end{equation}
Coupling constants flow down into the modified fixed relations and then 
follow the relations. 
Thus,  the evolution of the coupling constants toward the UV is the most
stable when the parameter satisfies the ``IR-fixed relations''.
The modified fixed relations are still almost the (${\cal N}$ = 2)-SUSY
relations, and hence the 1-loop evolution is almost correct for the
parameter satisfying the relations; beta functions are different from
those at one loop only by an order of $\alpha_{\rm GUT}$. 
Moreover, combinations such as $(\alpha_{\rm 2H}$ -- $\alpha_{\rm
2H}^{\lambda^{(')}})$ partially absorb the SU(5)$_{\rm GUT}$ 
contributions in the beta functions.   
Therefore, the corrections to the 1-loop evolution are estimated
conservatively from above when the SU(5)$_{\rm GUT}$ contribution is
purely added to the 1-loop beta functions, as we did in subsection
\ref{ssec:uncert2}.

\newpage

\begin{table}[ht]
\begin{center}
\begin{tabular}{c|ccccc}
Fields    & ${\bf 10}^{ij}$ & ${\bf 5}^*_i$ 
          & $X$ & $Q_i,\bar{Q}^i$ & $Q_6$,$\bar{Q}^6$ \\
\hline 
R charges & 1 & 1 & 2 & 0 & 0  \\
\end{tabular}
\caption{(Mod 4)-R charges of the fields in the 
SU(5)$_{\rm GUT}\times$U(2)$_{\rm H}$ model. }
\label{tab:u2-R}
\end{center}
\end{table}

\begin{table}[ht]
\begin{center}
\begin{tabular}{c|ccccccccc}
Fields &
 ${\bf 10}^{ij}$ & ${\bf 5}^*_i$ 
 & $H({\bf 5})^i$ & $\bar{H}({\bf 5}^*)_i $ 
 & $X^\alpha_{\;\;\beta}$ 
 & $Q^\alpha_{\;\; i}$ & $\bar{Q}^i_{\;\; \alpha}$ 
 & $Q^\alpha_6$ & $\bar{Q}^6_{\;\; \alpha}$  \\
\hline 
R charges &
 1 & 1 & 0 & 0 & 2 & $0$ & $0$ & $2$ & $-2$ \\
\end{tabular}
\caption{(Mod 4)-R charges of the fields in the 
SU(5)$_{\rm GUT}\times$U(3)$_{\rm H}$ model.}
\label{tab:u3-R4}
\end{center}
\end{table}

\begin{table}[ht]
\begin{center}
\begin{tabular}{ccccc}
\hline
({\bf 3},{\bf 2})$^{-\frac{5}{6}}$ & 
({\bf 1},{\bf 1})$^0$ &
({\bf 1},{\bf 1})$^0$  &
({\bf 1},{\bf adj.})$^0$ &
({\bf 1},{\bf adj.})$^0$ \\
\hline
m.vect. & m.vect. & $\chi + \chi^{\dagger}$ 
        & m.vect. & $\chi + \chi^{\dagger}$ \\
\hline
$M_G = $ & $M_{1V} =  $ & $M_{1C} =  $ &
 $M_{3V} =  $ & $M_{3C} =  $ \\
$\sqrt{2}g_{\rm GUT}v$     &
$\sqrt{2(g_{\rm 1H}^2 + 3 g_{\rm GUT}^2/5)} v $  &
$\sqrt{2}\lambda_{\rm 1H} v $ &
$\sqrt{2(g_{\rm 2H}^2 + g_{\rm GUT}^2)} v $  &
$\sqrt{2}\lambda_{\rm 2H} v $  \\
\hline
\end{tabular}
\caption{Summary of the particle spectrum around the GUT scale 
of the SU(5)$_{\rm GUT}\times$U(2)$_{\rm H}$ model.  
 The first line denotes the representation under the gauge group of the
MSSM. In the second line, m.vect. denotes ${\cal N}$ = 1 
 massive vector multiplets and $\chi + \chi^\dagger$ a pair of ${\cal N}$
 = 1 chiral and antichiral multiplets. 
In the last line, the mass of each multiplet are given 
in terms of gauge coupling constants 
and parameters in the superpotential (\ref{eq:super2}). 
}
\label{tab:u2-spec}
\end{center}
\end{table}

\begin{table}[ht]
\begin{center}
\begin{tabular}{ccccccc}
\hline
(${\bf 3},{\bf 2})^{-\frac{5}{6}}$ & 
$({\bf 3},{\bf 1})^{-\frac{1}{3}}$ &
$({\bf 3},{\bf 1})^{-\frac{1}{3}}$ &
$({\bf 1},{\bf 1})^0$ &
$({\bf 1},{\bf 1})^0$  &
$({\bf adj.},{\bf 1})^0$ &
$({\bf adj.},{\bf 1})^0$ \\
\hline
m.vect. & $\chi + \chi^{\dagger}$ & $\chi + \chi^{\dagger}$ &
m.vect. & $\chi + \chi^{\dagger}$ & m.vect. & $\chi + \chi^{\dagger}$ \\
\hline
$M_G = $ & $M_{H_c} =  $ & $M_{H_{\bar{c}}}=  $ &  $M_{1V} =  $ & $M_{1C} =  $ &
 $M_{8V} =  $ & $M_{8C} =  $ \\
$\sqrt{2}g v$     &
$h v $   &  $h' v$    &
$\sqrt{2(g_{\rm 1H}^2 + 2 g^2/5)} v $  &
$\sqrt{2}\lambda_{\rm 1H} v $ &
$\sqrt{2(g_{\rm 3H}^2 + g^2)} v $  &
$\sqrt{2}\lambda_{\rm 3H} v $  \\
\hline
\end{tabular}
\caption{Summary of the particle spectrum around the GUT scale 
of the SU(5)$_{\rm GUT}\times$U(3)$_{\rm H}$ model. 
The SU(5)$_{\rm GUT}$ gauge coupling  constant $g_{\rm GUT}$ is abbreviated
 as $g$ in this table. See the caption for Table \ref{tab:u2-spec} for
 the conventions in this table, replacing ``superpotential (\ref{eq:super2})"  
by ``superpotential (\ref{eq:super3})".}

\label{tab:U3-spec}
\end{center}
\end{table}


\end{document}